\documentclass[showpacs,amsmath,amssymb,twocolumn,superscriptaddress,notitlepage,preprintnumbers,pra]{revtex4-1}

\usepackage[dvips]{graphicx}
\usepackage{amsmath,amssymb,amsthm,mathrsfs,amsfonts,dsfont}
\usepackage[braket]{qcircuit}
\usepackage{dcolumn}
\usepackage{bm}
\usepackage{amsmath}
\usepackage{amssymb}
\usepackage{braket}
\usepackage{physics}
\usepackage{amsthm}
\usepackage{natbib}
\usepackage{mathtools}
\usepackage{diagbox}
\usepackage[utf8]{inputenc}
\usepackage[english]{babel}
\usepackage{scalerel}[2014/03/10]
\usepackage{stackengine}
\usepackage{algorithm}
\usepackage[noend]{algpseudocode} 
\usepackage{color}
\usepackage{orcidlink}
\usepackage{changes}

\usepackage{graphicx}

\usepackage[labelfont=bf]{subcaption}
\usepackage{multirow}

\usepackage{hyperref}
\hypersetup{colorlinks=true, linkcolor=blue, citecolor=blue, urlcolor=black }
\usepackage{sidecap}

\usepackage{qcircuit}

\newcommand{\melem}[3]{\langle #1|#2 |#3 \rangle}

\renewcommand{\qed}{\hfill\blacksquare}

\newcommand{\GG}{\mathcal{G}}

\newcommand{\bos}{\pmb}

\begin{document}

\title{Symmetry enhanced variational quantum imaginary time evolution}

\author{Xiaoyang Wang\orcidlink{0000-0002-2667-1879} }

\affiliation{School of Physics, Peking University, Beijing 100871, China
}
\affiliation{Collaborative Innovation Center of Quantum Matter, Beijing 100871, China}
\affiliation{Center for High Energy Physics, Peking University, Beijing 100871, China}

\author{Yahui Chai}
\affiliation{CQTA, Deutsches Elektronen-Synchrotron DESY, Platanenallee 6, 15738 Zeuthen, Germany}

\author{Maria Demidik}
\affiliation{CQTA, Deutsches Elektronen-Synchrotron DESY, Platanenallee 6, 15738 Zeuthen, Germany}
\affiliation{Computation-Based Science and Technology Research Center, The Cyprus Institute, 20 Kavafi Street, 2121 Nicosia, Cyprus}

\author{Xu Feng}
\affiliation{School of Physics, Peking University, Beijing 100871, China
}
\affiliation{Collaborative Innovation Center of Quantum Matter, Beijing 100871, China}
\affiliation{Center for High Energy Physics, Peking University, Beijing 100871, China}

\author{Karl Jansen}
\affiliation{CQTA, Deutsches Elektronen-Synchrotron DESY, Platanenallee 6, 15738 Zeuthen, Germany}
\affiliation{Computation-Based Science and Technology Research Center, The Cyprus Institute, 20 Kavafi Street, 2121 Nicosia, Cyprus}

\author{Cenk T\"uys\"uz\orcidlink{0000-0003-0257-9784}}
 \affiliation{CQTA, Deutsches Elektronen-Synchrotron DESY, Platanenallee 6, 15738 Zeuthen, Germany}
\affiliation{Institut für Physik, Humboldt-Universit\"at zu Berlin, Newtonstr. 15, 12489 Berlin, Germany}

\date{\today}

\begin{abstract}
    The variational quantum imaginary time evolution (VarQITE) algorithm is a near-term method to prepare the ground state and Gibbs state of Hamiltonians. Finding an appropriate parameterization of the quantum circuit is crucial to the success of VarQITE. This work provides guidance for constructing parameterized quantum circuits according to the locality and symmetries of the Hamiltonian. Our approach can be used to implement the unitary and anti-unitary symmetries of a quantum system, which significantly reduces the depth and degree of freedom of the parameterized quantum circuits. To benchmark the proposed parameterized quantum circuits, we carry out VarQITE experiments on statistical models. Numerical results confirm that the symmetry-enhanced circuits outperform the frequently-used parametrized circuits in the literature.
\end{abstract}

\maketitle
\section{Introduction}

The latest developments of quantum computers have perspectives in certain computational problems~\cite{Arute_2019,zhong_2020}. These experiments inspire many efforts to study the capabilities of the so-called \textit{noisy intermediate scale quantum} (NISQ) devices~\cite{Preskill_2018}. NISQ means the qubits are not protected by quantum error correction codes~\cite{Nielsen2000}. Thus, only short quantum circuits are available for computation. Variational quantum algorithms (VQAs) are NISQ algorithms utilizing short quantum circuits as subroutines in a hybrid quantum-classical optimization loop. This hybrid approach in VQAs is being widely used to deal with computationally expensive problems in quantum chemistry, statistical mechanics, particle physics and machine learning~\cite{McArdle_20, Klco_2018, Klco_2020, Ciavarella_2021, Funcke_2022, Clemente_strategies, Banuls2020, Bharti22}.

The effect of the optimization using VQAs is mainly determined by two factors. The first is the classical optimization algorithm, and the second is the parameterization of the quantum circuit, where the parameterized quantum circuit is usually called the ansatz. \textit{Variational quantum imaginary time evolution} (VarQITE) is a classical optimization algorithm that carries out quantum imaginary time evolution (QITE) to a system's initial state based on an ansatz~\cite{McArdle_19, Yuan_2019}. This algorithm is suitable for ground state and Gibbs state preparation if the ans\"atze have enough expressivity. Many different heuristically derived ans\"atze (e.g. unitary coupled cluster (UCC) ansatz~\cite{Peruzzo:2014} and Hamiltonian variational ansatz~\cite{farhi2014quantum, Wecker15}) exist in the literature~\cite{McArdle_20, TILLY20221}. However, these are in general designed specifically for the ground state preparation. In Ref.~\cite{Wang23}, the authors provide a systematic ansatz design strategy suitable for Gibbs state preparation, and the indication of the criticality of the classical Ising model can be observed by preparing Gibbs state using the ansatz. However, naively implementing the ansatz design strategy leads to deep quantum circuits, and the number of free parameters is too large to be optimized. In this article, we propose to use the locality and symmetries of a quantum system to improve the ansatz design strategy, so that much shallower ans\"atze with fewer free parameters can be constructed. Using numerical simulations, we confirm that the proposed ans\"atze have good expressivity, and outperform the standard hardware-efficient ans\"atze in Gibbs state preparation of statistical models.

Symmetries play a central role in various fields. The implementation of symmetries in the context of VQAs is widely discussed in the literature. Some explicit constructions of ans\"atze preserving particle number (which is a manifestation of the $U(1)$ symmetry in the system) and spin symmetry have been proposed in Ref.~\cite{Gard_20, Barron_21, Lyu2023symmetryenhanced}. Apart from ansatz construction, symmetries can be implemented in the measurement operation by post-processing~\cite{Bonet_18, Seki_20, Seki_22} or in the Hamiltonian construction by penalizing states of wrong symmetry~\cite{Ryabinkin_18, Kuroiwa_21}. The symmetry-enhanced QITE algorithm is found in Ref.~\cite{Sun_21}. A similar goal of our work has been proposed recently in the framework of geometric quantum machine learning(GQML)~\cite{Larocca_22, Meyer_22, Sauvage_22, Nguyen_22, Ragone_22}, though the goal is reached via a distinctly different route. In GQML, unitary symmetry transformations are used to simplify the existing ans\"atze, such as Hamiltonian variational ans\"atze for ground state preparation~\cite{Sauvage_22}, and the symmetries are identified in machine learning models. In this work, we propose new ans\"atze for Gibbs state preparation. We demonstrate the method on quantum chemistry and statistical models, where both unitary and anti-unitary symmetry transformations are considered.

In Section~\ref{sec:Preliminaries}, we review the strategy of ans\"atze design for Gibbs state preparation, and show how the symmetry can be incorporated in the ans\"atze using an example of Ising model. In Section~\ref{sec:Symmetries in quantum mechanics}, we show how the symmetry transformations can be represented using unitary or anti-unitary operators. We exhibit discrete symmetries for some statistical models and continuous symmetries for chemistry models. In Section~\ref{sec:Implementing symmetry in ansatz}, we present the framework for implementing symmetries in ans\"atze. We also show some examples of how the ans\"atze can be reduced for the statistical models and the chemistry models. To demonstrate the performance of the ans\"atze, we conduct numerical simulations in Section~\ref{sec:Numerical simulations}. We show that the ans\"atze before and after the symmetry reduction have the same ability in Gibbs state preparation, and both outperform the standard hardware-efficient ans\"atze.

\section{Ansatz design}\label{sec:Preliminaries}
For a quantum system with a Hamiltonian $H$ and an initial state $\ket{\psi}=\ket{\psi(0)}$, quantum imaginary time evolution (QITE) is designed to evolve the initial state according to 
\begin{align}
    \ket{\psi(\tau)}=\frac{e^{-\tau H}\ket{\psi}}{||e^{-\tau H}\ket{\psi}||},
    \label{eq:psi_QITE}
\end{align}
where $\tau$ is a real number denoting imaginary time, and the denominator $||e^{-\tau H}\ket{\psi}||=\sqrt{\bra{\psi}e^{-2\tau H}\ket{\psi}}$ is a normalization factor to guarantee the evolution’s unitarity. Due to the imaginary time $\tau$, Gibbs state of the quantum system can be prepared by choosing a specific initial state~\cite{Yuan_2019}, and the ground state is obtained when $\tau$ is large enough. 

The evolution in Eq.~(\ref{eq:psi_QITE}) can be performed on an ansatz by evolving its free parameters~\cite{McArdle_19, Yuan_2019}. However, due to the finite expressivity of the ansatz, usually, the evolution can not be carried out exactly. In this section, we introduce the strategy proposed in~\cite{Wang23} on constructing an ansatz to carry out the evolution with high accuracy. The strategy is based on the work by Motta et al. ~\cite{Motta_20}, which proposed a QITE algorithm using quantum circuits with deterministic rotation parameters. We call this algorithm \textit{deterministic quantum imaginary time evolution} (DetQITE) for convenience.

To implement the transformation in Eq.~(\ref{eq:psi_QITE}) on quantum computers, one can consider a Hermitian operator $\hat{A}$, such that the imaginary time evolved state can be written as 
\begin{align}
    \ket{\psi(\tau)}= e^{i \tau \hat{A}(\tau)}\ket{\psi},
    \label{eq:unitary-transformation}
\end{align}
where $\hat{A}(\tau)$ lives on the same Hilbert space with the Hamiltonian of the system. Assume the system is composed of $N_q$ qubits, then $\hat{A}(\tau)$ can be expanded using a complete Pauli-basis
\begin{align}
    \hat{A}(\tau)=\sum_i a_i(\tau) \sigma_i,
    \label{eq:A-definition}
\end{align}
where $\sigma_i\equiv \sigma_{\alpha_{1}}\ldots \sigma_{\alpha_{N_q}}$ is the tensor product of single-qubit Pauli operators $\sigma_{\alpha_{n}}=I,X,Y,Z$, corresponding to $\alpha_{n}=0,1,2,3$ at site $n$ (The tensor product notation $\otimes$ will be omitted through out this article). The tensor product Pauli operator $\sigma_i$ is sometimes called Pauli string. Its single-qubit element is called Pauli letter~\cite{Oliver_22}. The expansion coefficients $a_i$ are real due to the Hermicity of $\hat{A}(\tau)$ and the Pauli strings. In principle, one can construct a system that possesses Hamiltonian $\hat{A}(\tau)$ so that the unitary transformation in Eq.(\ref{eq:unitary-transformation}) can be realized naturally. However, practically, neither such a system nor the expansion coefficients $a_i$ are available. Instead, we consider Trotterizing the unitary evolution to several time slices
\begin{equation}
\begin{aligned}
    e^{i \tau \hat{A}(\tau)}&=(e^{-i\frac{\tau}{L}\sum_i a_i(\tau) \sigma_i})^L\\
    &=(\prod_i e^{-i\Delta\tau a_i \sigma_i})^L +\mathcal{O}(\Delta \tau^2),
    \label{eq:pauli-exponentials-expansion}
\end{aligned}
\end{equation}
where $L$ is the number of Trotter steps and $\Delta\tau\equiv \tau/L$. Then, a series of Pauli exponentials $e^{-i\theta \sigma_i}$ can realize the unitary imaginary propagator. In this work, we consider all these Pauli exponentials as the \textit{standard parameterized gate set}. The Pauli exponentials can be realized using quantum gates in a standard way. (See the textbook~\cite{Nielsen2000} for details. Here, we assume that the only available quantum gates are single-qubit Pauli rotation $R_X, R_Y, R_Z$ and two-qubit CNOT gates between arbitrary two qubits. All the notations follow the textbook~\cite{Nielsen2000}). 

According to Eq.~(\ref{eq:pauli-exponentials-expansion}), one can immediately realize that if one layer of the ansatz is constructed by Pauli exponentials with all the Pauli string on the system, and when the number of ansatz layers tends to infinity, this ansatz can carry out QITE for an arbitrary Hamiltonian. However, for a quantum system encoded on $N_q$ qubits, the number of Pauli strings is $4^{N_q}$. When $N_q$ is large, such an amount of Pauli exponentials is hard to experimentally realize on quantum devices, even if $L=1$. Thus, reductions are needed to the number of Pauli exponentials. Many frequently used ans\"atze have Pauli exponentials as the standard parameterized gate set, such as the chemically inspired UCC ans\"atze and Hamiltonian variational ans\"atze~\cite{McArdle_20}. These ans\"atze provide heuristic strategies to implement the reductions. In this article, reductions are implemented more systematically, which is inspired by DetQITE algorithm~\cite{Motta_20}. In the following two subsections, we demonstrate how to perform the reductions according to the locality and symmetries of the Hamiltonian.

\subsection{Reduction of Pauli exponentials using locality of Hamiltonian}
Consider the imaginary time evolution of a local interacting Hamiltonian
\begin{align}
    H=\sum_m H_m,
\end{align}
where $H_m$ is a local interaction term that acts on a local set of qubits, and the number of $H_m$ is polynomial as a function of the system size. Many physically relevant Hamiltonian can be written in this form. If the local interaction terms commute, one can realize the imaginary time propagator for each $H_m$ as in Eq.~(\ref{eq:unitary-transformation}). If the local interaction terms do not commute, one has to take Trotterization of the imaginary time propagator 
\begin{align}
    e^{-\tau H} = (\prod_m e^{-\Delta \tau H_m})^L+\mathcal{O}(\Delta \tau^2).
\end{align}
So that we only need to focus on the imaginary time evolution of each local interaction term $e^{-\Delta \tau H_m}$ where $\Delta \tau\equiv \tau/L$. Realizing this propagator on quantum computers means expanding this propagator on Pauli-basis as in Eq.~(\ref{eq:pauli-exponentials-expansion})
\begin{align}
    \frac{e^{-\Delta \tau H_m}\ket{\psi}}{||e^{-\Delta \tau H_m}\ket{\psi}||}=\prod_i e^{-i\Delta\tau a_i\sigma_i}\ket{\psi}+\mathcal{O}(\Delta \tau^2).
    \label{eq:local-interaction-expansion}
\end{align}
Strictly speaking, though the support of $H_m$ is small and local (support of a Pauli string is defined by the set of qubits on which the Pauli letters are not identity), the Pauli strings $\sigma_i$ still should have support on the whole lattice, so that the total number of Pauli exponentials $e^{-i\Delta\tau a_i\sigma_i}$ is $4^{N_q}$ for each local interaction term.

The total number of Pauli exponentials can be reduced from $4^{N_q}$ to a constant, if the correlation length of the system is finite. Specifically, as shown in Ref.~\cite{Motta_20}, when the correlation length of the system is finite, the right-hand side of Eq.~(\ref{eq:local-interaction-expansion}) can be well approximated by a part of Pauli exponentials among all the $4^{N_q}$ ones. These Pauli exponentials should have support \textit{constantly larger} than the support of $H_m$. The correlation length of the system is finite when the system is outside the critical region. Thus, the number of Pauli strings to be implemented for each local interaction term has no dependence on the system size, and the total number of Pauli exponentials $e^{-i\Delta\tau a_i\sigma_i}$ in the evolution circuits is a polynomial function of the system size, at least when the Hamiltonian is sufficiently far away from the critical point. In this article, we choose the support of Pauli strings equal to the support of $H_m$.

In DetQITE algorithm~\cite{Motta_20}, the expansion coefficients $a_i$ in Eq.~(\ref{eq:local-interaction-expansion}) are calculated by minimizing the square of the difference between the left-hand side and right-hand side of Eq.~(\ref{eq:local-interaction-expansion}). It leads to a linear system of equations
\begin{align}
    \sum_j M_{ij} a_j = V_i,
    \label{eq:DetQITE-linear-equation}
\end{align}
where 
\begin{equation}
\begin{aligned}
    M_{ij}&\equiv \Re(\bra{\psi}\sigma_i\sigma_j\ket{\psi}),\\
    V_i&\equiv \Im(\bra{\psi}\sigma_i H_m\ket{\psi}).
    \label{eq:M-V-calculation-measure}
\end{aligned}
\end{equation}
By solving this equation for each local interaction term and each layer, all the Pauli exponentials can be determined, and the quantum circuits are constructed following the right-hand side of Eq.~(\ref{eq:local-interaction-expansion}). 

DetQITE algorithm is computationally expensive. The number of Pauli exponentials after the locality reduction is still large, and the resulting quantum circuits have a large number of layers $L$. For these reasons, the algorithm is applicable for small quantum systems~\cite{Aydeniz_2020}. Thus, we further implement reductions using symmetries of the Hamiltonian and then convert the circuits into ans\"atze with small $L$.

\subsection{Reduction of Pauli exponentials using symmetries of Hamiltonian}\label{sec:cut-off using Hamiltonian's symmetry}

Using Eq.~(\ref{eq:DetQITE-linear-equation}), we can solve a priori that some of $a_j$ are zero, and thus the degree of freedom of the linear system can be simplified. The authors of Ref.~\cite{Sun_21} analyzed how the simplification can be achieved in the case of $\mathcal{Z}_2$ symmetry. We explain this idea using a concrete example of the Ising model. 

The Hamiltonian of Ising model is 
\begin{align}
    H_{\mathrm{Ising}}=-\sum_{\langle n,n' \rangle} Z_{n}Z_{n'},
    \label{eq:Ising-model}
\end{align}
where $\langle n,n' \rangle$ denotes the nearest neighbour coupling between the lattice sites $n$ and $n'$. $Z_{n}Z_{n'}$ is the local interaction term. As explained in the previous subsection, the imaginary time propagator $e^{-\Delta \tau Z_{n}Z_{n'}}$ can be approximated by $4^{2}=16$ Pauli exponentials on sites  $n$ and $n'$, which bring out $16$ $a_i$ to be solved by the linear system of equations. However, some of $a_i$ are strictly zero, as guaranteed by the system's symmetries. 

The Ising system has two symmetries. Firstly, all entries of $Z_{n}Z_{n'}$ under computational basis are purely real, thus when the initial state $\ket{\psi}$ is also a purely real wave function, the propagator $e^{-\Delta \tau H_m}/||e^{-\Delta \tau H_m}\ket{\psi}||$ should also be purely real. On the other hand, a Pauli string $\sigma_i$ is purely imaginary (real) if it consists of an odd (even) number of $Y$ letters, which corresponds to a purely real (imaginary) Pauli exponential $e^{-i\theta_i \sigma_i}$. In the next section, we will show that a purely real Hamiltonian corresponds to the \textit{time-reversal} symmetry. Thus, $10$ of $16$ Pauli strings with even numbers of $Y$ letters can be eliminated due to the time-reversal symmetry of the Ising model, and the corresponding $a_i$ are strictly zero. The left Pauli strings are $I_{n}Y_{n'},Y_{n}I_{n'},X_{n}Y_{n'},Y_{n}X_{n'},Z_{n}Y_{n'}$ and $Y_{n}Z_{n'}$. Secondly, the Ising system has the $\mathcal{Z}_2$ symmetry which is manifested by the commutation relation 
\begin{align}
    [\prod_{n} X_{n}, H_{\mathrm{Ising}}]=0,
    \label{eq:Ising-commutation-relation}
\end{align}
and we also require that the initial state $\ket{\psi}$ is invariant by the symmetry operator $\prod_{n} X_{n}$, i.e., 
\begin{align}
    \prod_{n} X_{n}\ket{\psi}=\ket{\psi}.
\end{align}
Then Ref.~\cite{Sun_21} shows that only Pauli strings commuting with $\prod_{n} X_{n}$ contribute to the imaginary propagator. So that only Pauli strings $Z_{n}Y_{n'}$ and $Y_{n}Z_{n'}$ have non-zero expansion coefficients. 

In summary, according to the symmetry analysis of the Ising model, we only need two Pauli exponentials to approximate the imaginary propagator of one local interaction term
\begin{align}
    \frac{e^{-\Delta \tau ZZ}\ket{\psi}}{||e^{-\Delta \tau ZZ}\ket{\psi}||}\simeq e^{-i\Delta\tau a_{ZY}ZY}e^{-i\Delta\tau a_{YZ}YZ}\ket{\psi}+\mathcal{O}(\Delta \tau^2).
    \label{eq:Ising-imaginary-propagator}
\end{align}
The number of Pauli exponentials is reduced compared with the one on the right-hand side of Eq.~(\ref{eq:local-interaction-expansion}).

\subsection{Ansatz design strategy and time evolution}\label{sec:ansatz construction and evolution}

All the above derivation requires a large number of layers, i.e., $L\sim\mathcal{O}(\tau/\Delta \tau)\gg 1$, while only shallow quantum circuits are available on NISQ devices. Ref.~\cite{Wang23} points out that a shallow ansatz can be designed such that its one layer is all the necessary Pauli exponentials presented in the previous subsections, and then convert all the rotation parameters in Pauli exponentials into free parameters. This ansatz needs the number of layers $L$ proportional to the system size, and has no explicit dependence on the imaginary time $\tau$. For the Ising model, the basic building block for each local interaction term in each layer is 
\begin{align}
    U_{l,nn'}(\theta_{l,nn'},\theta_{l,n'n})\equiv e^{-i\theta_{l,nn'} Z_{n}Y_{n'}}e^{-i\theta_{l,n'n} Y_{n}Z_{n'}},
\end{align}
where $\theta_{l,nn'}=\theta_{l,nn'}(\tau),\theta_{l,n'n}=\theta_{l,n'n}(\tau)$ are free parameters to be evolved as a function of the imaginary time $\tau$. $l$ is the layer index. Then the ansatz can be expressed by
\begin{align}
    \ket{\phi(\bos{\theta})}= \prod_{l=1}^{L} \prod_{\langle n,n' \rangle} U_{l,nn'}(\theta_{l,nn'},\theta_{l,n'n})\ket{\psi},
\end{align}
where $L$ is the number of ansatz layers. In this expression, since $U_{l,nn'}$ on different sites $n,n'$ may not commute, different orders of $U_{l,nn'}$ in the product $\prod_{\langle n,n' \rangle}$ lead to different ans\"atze, and have impacts on the ans\"atze expressivity. For Gibbs state preparation of the Ising model, the ladder-arranged circuit is recommended~\cite{Wang23}. We will take this arrangement in the numerical simulations of this work.

Using this ansatz, the imaginary time evolution of $\ket{\psi}$ can be translated to the evolution of free parameters $\bos{\theta}(\tau)$. In VarQITE algorithm~\cite{McArdle_19, Yuan_2019}, the evolution of $\bos{\theta}(\tau)$ can be derived using McLachlan variational principle, where another linear system of equations is constructed 
\begin{align}
    \sum_j \widetilde{M}_{ij} \dot{\theta}_j = \widetilde{V}_i ,
    \label{eq:VarQITE-linear-equation}
\end{align}
with
\begin{equation}
\begin{aligned}
    \widetilde{M}_{ij}&\equiv \Re\left[\frac{\partial \bra{\phi(\bos{\theta})}}{\partial \theta_{i}}\frac{\partial \ket{\phi(\bos{\theta})}}{\partial \theta_{j}}\right],\\
    \widetilde{V}_{i}&\equiv -\Re\left[\frac{\partial \bra{\phi(\bos{\theta})}}{\partial \theta_{i}} H\ket{\phi(\bos{\theta})}\right].
    \label{eq:M-V-definition}
\end{aligned}
\end{equation}
Here $\widetilde{M}$ is a matrix, and $\widetilde{V}$ is a vector. They have the same dimensions as the number of free parameters, and both can be evaluated on quantum computers using Hadamard test~\cite{McArdle_19, Funcke2021dimensional}. Then by solving the linear system $\dot{\bos{\theta}}=\widetilde{M}^{-1}\widetilde{V}$ using pseudo-inverse~\cite{McArdle_19,kress2012numerical}, the evolution of $\bos{\theta}(\tau)$ can be determined using Euler method
\begin{align}
    \bos{\theta}(\tau_0+\delta \tau) \simeq \bos{\theta}(\tau_0)+\dot{\bos{\theta}}\delta\tau,
    \label{eq:Euler-integration}
\end{align}
where $\delta\tau$ is the Euler step length.

The advantage of the proposed ansatz is that the whole process of imaginary time evolution can be performed on it, as long as the layer is large enough and the system is far from the critical point, guaranteed by the DetQITE algorithm. It is not the case for UCC or Hamiltonian variational ansatz. Thus, the proposed ansatz is suitable for Gibbs state preparation. The expressivity of the ansatz can be systematically improved by increasing the number of layers and the support of the Pauli exponentials.

\section{Symmetries in quantum mechanics}\label{sec:Symmetries in quantum mechanics}
 
Symmetries in quantum mechanics are captured by \textit{groups}. For example, in the Ising model, all the computational basis states are the system's energy eigenstate. The operation of $0\leftrightarrow 1$ flip (denoted as $g$) applying on a computational basis state does not change the state energy. Together with the identity operation $e$, the symmetry group of the Ising model contains two elements 
\begin{align}
    \GG_{\mathrm{Ising}}=\{e,g\}.
\end{align}
A more compact description of a group is by its \textit{generators}. Each element in the group can be expressed as a multiplication of the group generators. In the Ising model, the element $g$ is the generator of the Ising symmetry group, and the symmetry group can also be presented by
\begin{align}
    \GG_{\mathrm{Ising}}=\langle g\rangle.
\end{align}
Since the cardinality of the generating set is usually much less than that of the whole group, we use generating sets to present groups in the following contents.

In the Ising model, the flip operation $g$ can be represented by the matrix
\begin{align}
    U_g=\prod_{n} X_{n},
    \label{eq:Ising-symmetry-operation}
\end{align}
which is unitary, i.e., $U_gU_g^{\dagger}=I$. This fact can be generalized to all the symmetry operations of quantum mechanical systems, as depicted by Wigner's theorem~\cite{wigner_1959}, which states that symmetry transformation is represented as a unitary or an anti-unitary operator in the system's Hilbert space.

Wigner's theorem guides the classification of symmetry operations of a quantum system: unitary operator and anti-unitary operator. The unitary operator can be further classified into discrete symmetry and continuous symmetry. The anti-unitary operator is commonly used in the time-reversal transformation, which can only be discrete. In the following subsections, we discuss these symmetries in detail and show some models possessing these symmetries. 

\subsection{Discrete symmetry}\label{sec:Discrete symmetry}
Discrete symmetry is characterized by a discrete group that contains a finite number of elements. In this work, we focus on statistical models possessing global permutation symmetry, global additive symmetry and local additive symmetry, corresponding to the Potts model, clock model and $\mathcal{Z}_Q$ gauge model. With the VarQITE algorithm, we can generate Gibbs states of these statistical models on quantum computers and study their critical behaviours~\cite{Wang23}. We will also discuss how to encode these models on quantum computers.

The Hamiltonian of the $Q$-state Potts model~\cite{Wu_82} reads
\begin{align}
    H_{\mathrm{Potts}}=-\sum_{\langle n,n'\rangle}\delta(k_{n},k_{n'}),
    \label{eq:potts-hamiltonian}
\end{align}
where $k_{n},k_{n'} \in \{0,1,\ldots, Q-1\}$ are labels of states. $n,n'$ denote sites on a $2d$ square lattice and $\langle n,n'\rangle$ denotes nearest neighbor coupling. The $\delta$ function represents that the system energy will lower if the nearest neighbour two sites are in the same state. The Ising model in Eq.~(\ref{eq:Ising-model}) is a specific $Q$-state Potts model corresponding to $Q=2$. Similar to the Ising model, the system energy of the Potts model is unchanged by arbitrarily permuting the labels of the states. Thus, the symmetry group of the $Q$-state Potts model is the permutation group $\mathcal{S}_Q$, of which elements map $\{0,1,\ldots, Q-1\}$ to its arbitrary rearrangement. Thus, the $\mathcal{S}_Q$ group has $Q!$ elements. $\mathcal{S}_Q$ group can be generated by elementary operations
\begin{align}
    \mathcal{S}_Q=\langle (0,1),(1,2)\ldots (Q-1,0)\rangle.
\end{align}
The notation $(0,1)$ means interchanging labels $0$ and $1$, $(1,2)$ means interchanging labels $1$ and $2$; and so on. The number of generators of $\mathcal{S}_Q$ group is $Q$, which is much less than the total number of group elements.

To encode the Potts model on quantum computers, we choose the binary encoding, i.e., we use $P$ qubits encoding one spatial site of the $Q=2^P$-state Potts model. Thus, the Potts Hamiltonian Eq.~(\ref{eq:potts-hamiltonian}) has computational basis vectors as its eigenstates, and it can be written as a linear combination of Pauli-$Z$ strings(The Pauli string contains only $Z$ and $I$ letters). The symmetry operation can also be represented under the computational basis. More details on the representation of the Potts model and the symmetry are shown in Appendix~\ref{app:Statistical models and their representation}.

Replacing the $\delta$ function in Potts Hamiltonian with the cosine function leads to the $Q$-state clock model~\cite{Wu_82}
\begin{align}
    H_{\mathrm{clock}} = -\sum_{\langle n,n'\rangle}\cos (\theta_{n}-\theta_{n'}),
    \label{eq:clock-model}
\end{align}
where $\theta_{n}\equiv 2\pi k_{n}/Q, k_{n} \in \{0,1,\ldots, Q-1\}$. $\theta_{n}, \theta_{n'}$ can be interpreted as angles taking equal-spaced values from interval $[0,2\pi)$. Thus, the angles difference $(\theta_{n}-\theta_{n'})$ determines the energy of the nearest neighbour coupling $\langle n,n'\rangle$, while the absolute values of the angles are not important. It indicates that the clock model is invariant by rotating all the angles $\theta_{n}$ within a plane. These discrete rotation operations form the $\mathcal{Z}_Q$ group, which denotes the additive group of integers modulo $Q$. The elements of $\mathcal{Z}_Q$ map integers $\{0,1,\ldots, Q-1\}$ to the ones adding a constant(modulo $Q$). Thus, there are $Q$ elements in the group, and the set of $\mathcal{Z}_Q$ operations is a subset of $\mathcal{S}_Q$(It means the additive group is a \textit{subgroup} of the permutation group $\mathcal{S}_Q$). The generating set of $\mathcal{Z}_Q$ contains only one permutation
\begin{align}
    \mathcal{Z}_Q=\langle (0,1,\ldots, Q-1)\rangle.
\end{align}
The notation $(0,1,\ldots, Q-1)$ means a cyclic permutation of label $0\rightarrow 1,1\rightarrow 2,\ldots$ and $Q-1\rightarrow 0$.

The encoding of the clock model under the computational basis is similar to the Potts model. The resulting Hamiltonian contains only Pauli-$Z$ strings, and the symmetry operation can be encoded using the computational basis. See more details in Appendix~\ref{app:Statistical models and their representation}.

The symmetric operations introduced above are label-changing operations on \textit{all} sites. Some statistical models are invariant by operations on \textit{local} sites. We call the $Q$-state Potts model, $Q$-state clock model possessing \textit{global} $\mathcal{S}_Q$, \textit{global} $\mathcal{Z}_Q$ symmetry. In the following, we introduce a model possessing \textit{local} $\mathcal{Z}_Q$ symmetry, which is named $\mathcal{Z}_Q$ \textit{gauge model}.

\begin{figure}
    \centering
    \includegraphics[width=0.49\textwidth]{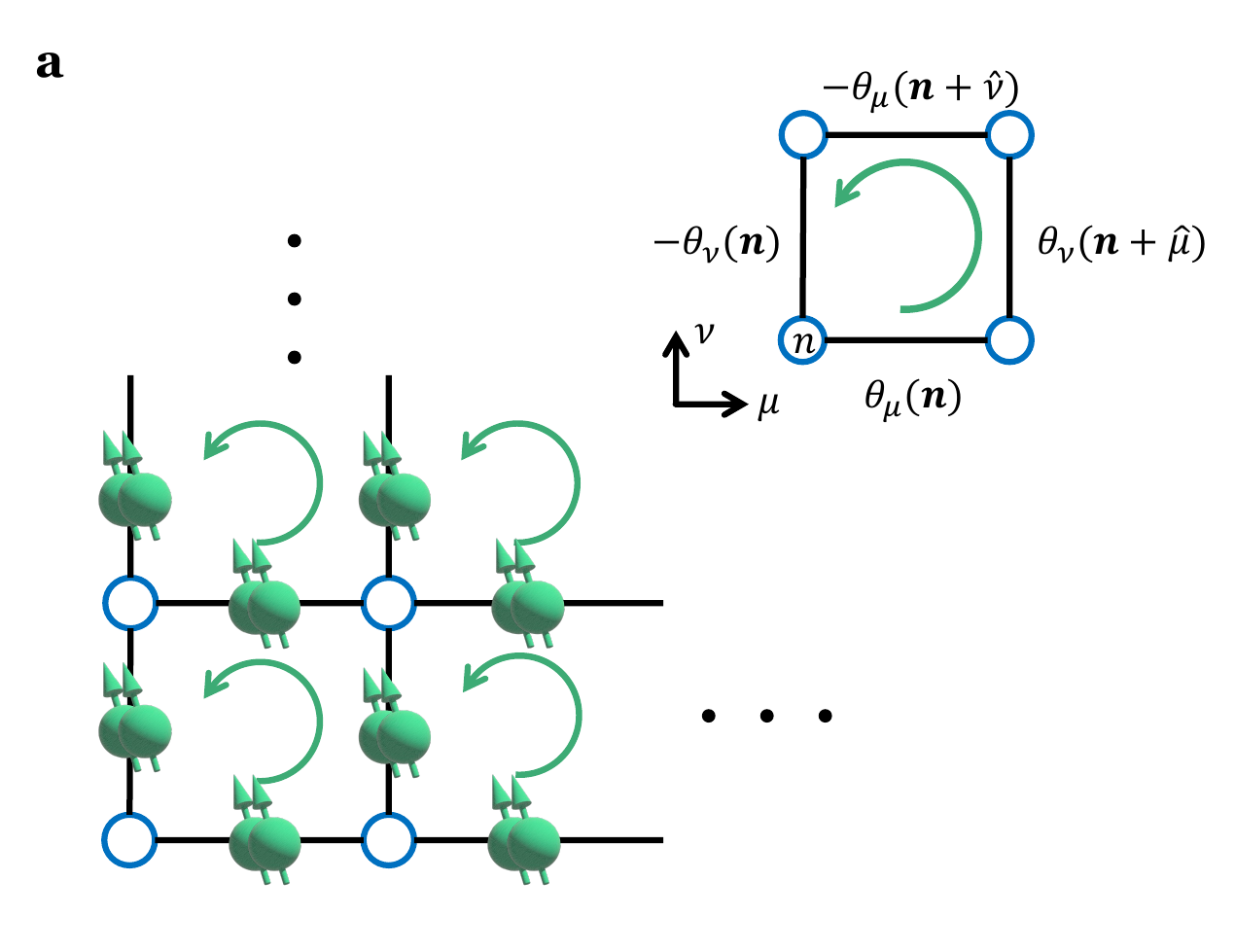}
    \includegraphics[width=0.49\textwidth]{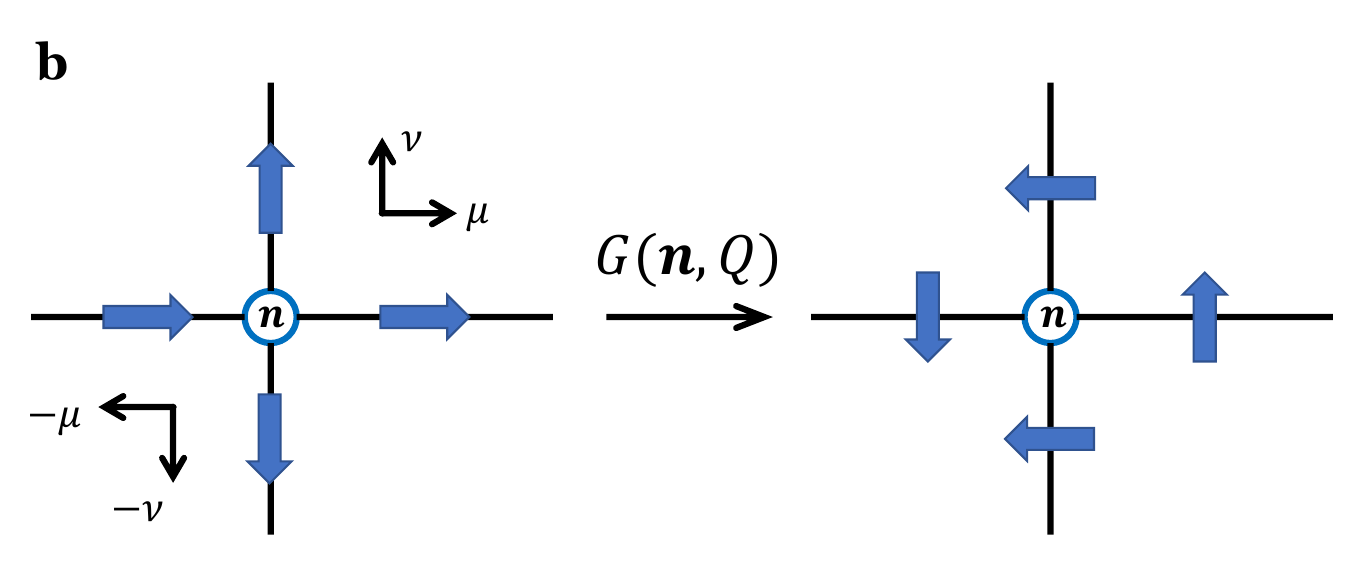}
    \caption{(\textbf{a}) Illustration of the $\mathcal{Z}_Q$ gauge model on $2d$ square lattice and the layout of qubits. The subplot on the upper right corner demonstrates one plaquette of the $\mathcal{Z}_Q$ gauge model.(\textbf{b}) Local gauge transformation in $\mathcal{Z}_4$ gauge model. The orientations of the blue arrow denote values of the angles $\theta_{\pm \mu/\nu}(n)$. The arrows on the positively oriented (up and right) links rotate anti-clockwise for $90^{\circ}$ by $G(n,4)$, while the arrows on the negatively oriented (down and left) links rotate clockwise for $90^{\circ}$. This transformation leaves Hamiltonian $H_{\mathrm{gauge}}$ unchanged.}
    \label{fig:plaquette}
\end{figure}

The Hamiltonian of the $\mathcal{Z}_Q$ gauge model~\cite{KogutRMP_79} reads
\begin{equation}
\begin{aligned}
    H_{\mathrm{gauge}} = -\sum_{n,\nu>\mu}\cos[&\theta_{\mu}(n)+\theta_{\nu}(n+\hat{\mu}) \\ 
    & -\theta_{\mu}(n+\hat{\nu})-\theta_{\nu}(n) ],
    \label{eq:z-q-gauge-model-hamiltonian}
\end{aligned}
\end{equation}
where $\theta_{\mu}(n)\equiv 2\pi k_{\mu}(n)/Q, k_{\mu}(n) \in \{0,1,\ldots, Q-1\}$. The angle defined here is similar to the one defined in the clock model. The difference is that here, $\mu$ denotes the link connecting two nearest neighbour sites starting from $n$ to $n+\hat{\mu}$ ($\hat{\mu}$ is a unit vector aligned with the link $\mu$), so that $\theta_{\mu}(n)$ denotes an angle living on a link $n\rightarrow n+\hat{\mu}$, but not on a node as in the clock model. The local interaction term of Eq.~(\ref{eq:z-q-gauge-model-hamiltonian}) is a real cosine function that is usually called \textit{plaquette}. An illustration of the $\mathcal{Z}_Q$ gauge model on $2d$ square lattice and a plaquette are shown in figure~\ref{fig:plaquette}\textbf{a}. To encode this model on qubits, we use a quantum register including several qubits to represent one angle on a link.

$\mathcal{Z}_Q$ gauge model preserves the local $\mathcal{Z}_Q$ symmetry, i.e., the system energy is invariant by a local rotating operation $G(n, Q)$, which rotates the angles a constant $2\pi/Q$ on the links emitted from the site $n$. The rotation is anti-clockwise for the positively oriented links $\mu$ and $\nu$ and clockwise for the negatively oriented links $-\mu$ and $-\nu$, as shown in Figure~\ref{fig:plaquette}\textbf{b}. The figure presents the local rotation in the $\mathcal{Z}_4$ gauge model. The state of each link is represented by an arrow oriented to $Q=4$ directions. The Hamiltonian Eq.~(\ref{eq:z-q-gauge-model-hamiltonian}) is invariant by the transformation $G(n,Q)$. Because all the plaquettes adjacent to site $n$ have two angles changed by $G(n,Q)$, and their changes cancel in the cosine function in all plaquettes adjacent to $G(n, Q)$. Local rotating operations on all the sites of lattice $\Lambda$ form a group that is generated by 
\begin{align}
    \mathcal{Z}_{Q,\mathrm{local}}=\langle \{G(n,Q)|n\in\Lambda\}\rangle.
    \label{eq:z-q-local}
\end{align}

One can observe different phase transitions and critical behaviours by studying the thermal properties of the statistical models introduced above. For example, the $Q=2$ Potts model, i.e., the Ising model, has second-order phase transition~\cite{Onsager1944, Schultz1964}, while the Potts model with large $Q$ has first-order phase transition~\cite{Wu_82}. The continuous limit $Q\rightarrow\infty$ of $Q$-state clock model corresponds to XY-model, which exhibits the Kosterlitz-Thouless transition~\cite{Kosterlitz_1973}. In this article, we study how to encode the internal symmetries of the statistical models on variational ans\"atze and use VarQITE to investigate their thermal properties. 

\subsection{Continuous symmetry}\label{sec:Continuous symmetry}

The continuous limit $Q\rightarrow\infty$ of the global $\mathcal{Z}_Q$ symmetry leads to the global $U(1)$ symmetry. The global $U(1)$ symmetry guarantees the conservation of electron charge numbers in condensed matter physics, quantum chemistry and materials science. The second-quantized form~\cite{McArdle_20, Shi_2021} of Hamiltonians in these systems can be formally written as
\begin{align}
H_{\mathrm{chem}}=\sum_{ij}h_{ij}\hat{a}_i^{\dagger}\hat{a}_j+\sum_{ijkl}V_{ijkl}\hat{a}_i^{\dagger}\hat{a}_j^{\dagger}\hat{a}_k\hat{a}_l.
\label{eq:quantum-chemistry-model}
\end{align}
Here we use $ijkl$ to index an appropriate chemical basis, such as the basis introduced in Ref.~\cite{McArdle_20}. $h_{ij},V_{ijkl}$ are some complex numbers and $\hat{a}_i^{(\dagger)}$ is the fermion annihilation (creation) operator. Global $U(1)$ symmetry means that the Hamiltonian is invariant by the transformation $g_{\alpha}$
\begin{align}
    \hat{a}_i^{\dagger}\stackrel{g_{\alpha}}{\longrightarrow}e^{i\alpha}\hat{a}_i^{\dagger},\quad \hat{a}_i\stackrel{g_{\alpha}}{\longrightarrow}e^{-i\alpha}\hat{a}_i,
    \label{eq:U1-transform}
\end{align}
where $\alpha$ is an arbitrary real number. We see that the chemical Hamiltonian is invariant because every local interaction term in the Hamiltonian has an equal number of creation and annihilation operators.

One can map the fermion operators onto qubits using encoding methods such as Jordan-Wigner (J-W) transformation~\cite{Jordan_28, Banks1976, OpenFermion}, and the representation of the $U(1)$ transformation Eq.~(\ref{eq:U1-transform}) on a quantum computational basis can be derived. For example, with J-W transformation~\cite{OpenFermion}, we have
\begin{equation}
\begin{aligned}
&\ket{0011}=\hat{a}_0^{\dagger}\hat{a}_1^{\dagger}\ket{0000} \\
    \stackrel{g_{\alpha}}{\longrightarrow} &e^{2i\alpha}\hat{a}_0^{\dagger}\hat{a}_1^{\dagger}\ket{0000}=e^{2i\alpha}\ket{0011}.
    \label{eq:U(1) on hartree fock state}
\end{aligned}
\end{equation}
Thus, the $U(1)$ transformation can be represented using a diagonal unitary matrix on the computational basis
\begin{align}
    (U_{g_{\alpha}}^{\operatorname{J-W}})_{xy}=\delta_{x,y} e^{i|x|\alpha},
    \label{eq:U(1)-representation}
\end{align}
where $x,y$ are bit strings and $|x|$ is the Hamming weight of string $x$~\cite{Nielsen2000}.

\subsection{Time-reversal symmetry}

Time-reversal (TR) symmetry is distinguished from symmetries introduced above, since it is represented using an anti-unitary operator. In this subsection, we first introduce how to represent anti-unitary operators mathematically. Then we define the TR invariance of a system and present a Hamiltonian possessing TR symmetry.

An anti-unitary operator can be decomposed by 
\begin{align}
    \mathcal{T}=UK,
    \label{eq:anti-unitary operator}
\end{align}
where $U$ is a unitary operator and $K$ is a complex conjugation operator. The complex conjugation operator makes $\mathcal{T}$ anti-unitary because its action on the operator and inner product is like
\begin{equation}
\begin{aligned}
    KOK^{-1}&=KOK=\overline{O}, \\
    \bra{\phi}K\ket{\psi}&=\bra{\phi}K^{-1}\ket{\psi} = \overline{\bra{K\phi}\ket{\psi}},
    \label{eq:K-effect}
\end{aligned}
\end{equation}
where the overline $\overline{(\cdot)}$ denotes complex conjugation, and $\overline{O}$ means complex conjugation on each entry of $O$. These actions will be utilized in the later derivations. The actions of $K$ are distinguished from those of unitary operators. For example, the unitary operator $U$ acting on an inner product is 
\begin{align}
    \bra{\phi}U^{-1}\ket{\psi} = \bra{U\phi}\ket{\psi},
\end{align}
with no overline on the right-hand side, which differs from the one in the second row of Eq.~(\ref{eq:K-effect}).

In numerical computation, we say a system is TR invariant if the system Hamiltonian is purely real, i.e., the Hamiltonian commutes with the complex conjugation operator $K$
\begin{align}
    \overline{H}=H\Leftrightarrow [K, H]=0.
    \label{eq:TR-invariance}
\end{align}
Instead of using a general anti-unitary operator $\mathcal{T}=UK$, only $K$ remains. It is because $U$ has been assumed to be absorbed in the Hamiltonian $H$, where $U$ behaves like a representation transformation.

Sometimes a system is physically TR invariant, but the Hamiltonian of the system is not purely real. In that case, one can find an appropriate representation to make the Hamiltonian real. For example, assume the Hamiltonian consists of nearest neighbour fermionic hopping operators
\begin{align}
    H_{\mathrm{hop}}=\sum_i t_{i,i+1},
    \label{eq:hopping-Hamiltonian}
\end{align}
where $t_{i,i+1}\equiv \hat{a}_{i}^{\dagger}\hat{a}_{i+1}+\hat{a}_{i+1}^{\dagger}\hat{a}_{i}$. This Hamiltonian is physically TR invariant and can be written as a matrix by J-W transformation. One way of taking J-W transformation~\cite{Banks1976} is
\begin{align}
    t_{i,i+1}\rightarrow \frac{1}{2}(Y_{i}X_{i+1}-X_{i}Y_{i+1}).
    \label{eq:ti-JW-encoding-1}
\end{align}
Because Pauli letter $X$ is purely real and $Y$ is purely imaginary, we have
\begin{equation}
\begin{aligned}
    K(Y_{i}X_{i+1}-X_{i}Y_{i+1})K^{-1}&=\overline{Y}_{i}\overline{X}_{i+1}-\overline{X}_{i}\overline{Y}_{i+1}\\
    &=-(Y_{i}X_{i+1}-X_{i}Y_{i+1}).
\end{aligned}
\end{equation}
Thus the resulting matrix of the Hamiltonian is not purely real. Another way of taking J-W transformation~\cite{OpenFermion} is
\begin{align}
    t_{i,i+1}\rightarrow \frac{1}{2}(X_{i}X_{i+1}+Y_{i}Y_{i+1}).
    \label{eq:ti-JW-encoding-2}
\end{align}
This matrix is different from the one in Eq.~(\ref{eq:ti-JW-encoding-1}) by a representation transformation, but is purely real. Thus, one should adopt the second J-W transformation to manifest the TR invariance of the Hamiltonian in Eq.~(\ref{eq:hopping-Hamiltonian}).

Making Hamiltonian purely real is beneficial to numerical computation. In classical computation, half of the memory can be saved if one wants to store all the entries of a Hamiltonian. On the quantum side, we will see that the ansatz is much shallower if the system Hamiltonian is purely real.

\section{Implementing symmetry in ans\"atze}\label{sec:Implementing symmetry in ansatz}

This section presents the general formalism to implement symmetry in the construction of variational ans\"atze. We have discussed how to implement symmetry to ans\"atze for the Ising model in section \ref{sec:cut-off using Hamiltonian's symmetry}, and here, we apply that idea to more general scenarios. 

Remember that we aim to constrain the expansion coefficients vector $\bos{a}$ with entries $(\bos{a})_i=a_i$ using a symmetry group $\GG$. The expansion coefficients are given by solving the linear system of equations~\cite{Motta_20}
\begin{align}
    M \bos{a} = V,
    \label{eq:DetQITE-linear-equation-again}
\end{align}
where the elements of the matrix $M$ and vector $V$ are given by Eq.~(\ref{eq:M-V-calculation-measure}). Assume that $g_{\bos{\alpha}}$ is an element of the symmetry group $\GG$ and $\mathcal{U}_{g_{\bos{\alpha}}}$ is its representation, which could be unitary or anti-unitary. $\bos{\alpha}$ is a vector of group parameters. We assume that $\mathcal{U}_{g_{\bos{\alpha}}}$ possesses the following three properties
\begin{itemize}
    \item \textbf{Commutation relation} 
    \begin{align}
        [\mathcal{U}_{g_{\bos{\alpha}}}, H_m]=0.
        \label{eq:Commutation relation}
    \end{align}
    \item \textbf{Invariance of the initial state} 
    \begin{align}
    \ket{\mathcal{U}_{g_{\bos{\alpha}}}\psi}=e^{if(\bos{\alpha})}\ket{\psi}, \bra{\mathcal{U}_{g_{\bos{\alpha}}}\psi}=\bra{\psi}e^{-if(\bos{\alpha})},
    \label{eq:Invariance of the initial state}
    \end{align}
    where $f(\bos{\alpha})$ is an arbitrary scalar function.
    \item \textbf{Transformation in Pauli-basis}
    \begin{align}
        \mathcal{U}_{g_{\bos{\alpha}}}^{-1} \sigma_i \mathcal{U}_{g_{\bos{\alpha}}}=\sum_{j} c^{(\bos{\alpha})}_{ij} \sigma_j,
        \label{eq:Transformation in Pauli-basis}
    \end{align}
    where $c^{(\bos{\alpha})}_{ij}$ are expansion coefficients in Pauli-basis.
\end{itemize}
The commutation relation in Eq.~(\ref{eq:Commutation relation}) is the intrinsic property of the symmetry transformation, as we have seen in the previous section. Invariance of the initial state in Eq.~(\ref{eq:Invariance of the initial state}) is a requirement on the initial state of QITE. This requirement is natural as the initial state usually keeps a settled quantum number before the evolution, so that the state is invariant by the symmetry transformation up to a global phase $e^{if(\bos{\alpha})}$ (See Eq.~(\ref{eq:U(1) on hartree fock state}) as an example of the global phase). The third property in Eq.~(\ref{eq:Transformation in Pauli-basis}) characterizes how the symmetry transformation behaves on Pauli-basis. It leads to the definition of $c^{(\bos{\alpha})}$ matrix, whose existence is guaranteed by the completeness of the Pauli-basis. $c^{(\bos{\alpha})}$ matrix plays a central role in our following derivations. In Appendix~\ref{app:cm-matrix}, we show that all entries in $c^{(\bos{\alpha})}$ matrix are real  and it is a real orthogonal matrix 
\begin{align}
    [c^{(\bos{\alpha})}]^{-1}=[c^{(\bos{\alpha})}]^T,
\end{align}
where $(\cdot)^T$ denotes matrix transpose.

One has to notice that if matrix $M$ is not full rank, the solution of the linear system Eq.~(\ref{eq:DetQITE-linear-equation-again}) is not unique. Explicitly, because $M$ matrix is a real symmetric matrix~\cite{Motta_20}, it has spectrum decomposition
\begin{align}
    M=\sum_{j=1}^r d_j \ket{j}\bra{j},
\end{align}
where $d_j$ is the eigenvalue, $\ket{j}$ is the eigenvector and $r$ is the rank of $M$. Here, we use $\ket{\cdot}$ and $\bra{\cdot}$ to represent a column vector and its dual, but have no meaning of quantum states. $r$ is the rank of the matrix, and the solution of the linear system Eq.~(\ref{eq:DetQITE-linear-equation-again}) is not unique if $r$ is smaller than the dimension of the $M$ matrix. However, to carry out DetQITE, we only need to choose a special solution of $\bos{a}$. We assume that the special solution is provided by the \textit{pseudo-inverse}~\cite{kress2012numerical}
\begin{align}
    \bos{a}_0 \equiv \sum_{j=1}^r \frac{1}{d_j}\bra{j}\ket{V}\ket{j}.
    \label{eq:a0-pseudo-inverse}
\end{align}

With the above preparation, we can see how a symmetry transformation constrains the expansion coefficients $\bos{a}_0$, as shown in the following theorem. \\
\textbf{Theorem}. Given that the expansion coefficients $\bos{a}_0$ solved by pseudo-inverse of the linear system $\sum_j M_{ij} a_j = V_i$, and the symmetry group $\GG$, of which element $g_{\alpha}$ possesses properties in Eq.~(\ref{eq:Commutation relation})-(\ref{eq:Transformation in Pauli-basis}). Then coefficients $\bos{a}_0$ satisfy the following constraints
\begin{align}
    c^{(\bos{\alpha})}\bos{a}_0=\xi \bos{a}_0,\quad \forall g_{\bos{\alpha}}\in \GG,
    \label{eq:master-formula}
\end{align}
with $\xi=+1$ for unitary symmetry transformation and $\xi=-1$ for anti-unitary symmetry transformation.

Eq.~(\ref{eq:master-formula}) is the main formula to implement symmetry constraints in the ans\"atze carrying out QITE, where each element $g_{\bos{\alpha}}$ in $\GG$ provides a constraint to $\bos{a}_0$. To fully consider the symmetry constraints, one should calculate $c^{(\bos{\alpha})}$ matrix and solve Eq.~(\ref{eq:master-formula}) for every $g_{\bos{\alpha}}\in \GG$. However, considering a discrete group, as discussed in the previous section, sometimes the number of elements in the group is large, and the above calculation consumes a lot of computational resources. The calculation procedure can be simplified if only the constraints from the generating set need to be considered, since the group's generating set is typically much smaller than the whole group. This simplification is feasible, as presented in the following corollary. Here we focus on the unitary symmetry transformations, where $\xi=1$ and
\begin{align}
    c^{(\bos{\alpha})}\bos{a}_0=\bos{a}_0,\quad \forall g_{\bos{\alpha}}\in \GG.
    \label{eq:master-formula-unitary}
\end{align}
\textbf{Corollary}. Solving constraints in Eq.~(\ref{eq:master-formula-unitary}) for all the elements $g_{\bos{\alpha}}\in \GG$ is equivalent to considering elements from the generating set of $\GG$.

By considering elements from the generating set only, the number of constraints to be solved is reduced. Proofs of the theorem and the corollary are provided in Appendix~\ref{app:theorem-proof}. In Appendix~\ref{app:theorem-proof}, we also discuss the relationship of Eq.~(\ref{eq:master-formula-unitary}) to the main formula used in the framework of GQML. In the following subsections, we provide examples to explicitly solve the constraints in Eq.~(\ref{eq:master-formula})

\subsection{Example on discrete symmetry}

We start with the Ising model 
\begin{align}
    H_{\mathrm{Ising}}=-\sum_{\langle n,n' \rangle} Z_{n}Z_{n'},
    \label{eq:Ising-model-on-symmetry}
\end{align}
which possesses the discrete symmetry represented by $U_g = \prod_{n} X_{n}$. We check the three properties Eq.~(\ref{eq:Commutation relation})-(\ref{eq:Transformation in Pauli-basis}) of the symmetry transformation. First, the commutation relation of each local interaction term is
\begin{align}
    [U_g, ZZ]=0.
\end{align}
Second, the initial state we choose for VarQITE is
\begin{align}
    \ket{\psi}=\ket{++\ldots +}\equiv \ket{\tilde{+}},
    \label{eq:Ising-initial-state}
\end{align}
which is invariant by the action of symmetry transformation $U_g$, thus Eq.~(\ref{eq:Invariance of the initial state}) is satisfied. Finally, we calculate the $c$ matrix restricted on two qubits according to Eq.~(\ref{eq:Transformation in Pauli-basis}), where $U_g=XX$, and we find
\begin{align}
    c_{ij}=\left\{\begin{array}{rc}
        \delta_{i,j},& \mathrm{for}~[\sigma_i,XX]=0; \\
        -\delta_{i,j} ,& \mathrm{for}~ \{\sigma_i,XX\}=0,
    \end{array}\right.
    \label{eq:c-for-pauli-symmetry}
\end{align}
We see that the $c$ matrix is diagonal and its entries have different values if the associated $\sigma_i$ commute or anti-commute with $XX$. Taking this matrix into Eq.(\ref{eq:master-formula-unitary}), we find that the coefficient $(\bos{a}_0)_i$ vanishes, if the corresponding $\sigma_i$ anti-commute with $XX$, i.e., 
\begin{align}
    (\bos{a}_0)_i=0,~  \forall \sigma_i:~\{\sigma_i,XX\}=0.
\end{align}
Thus, we can discard the corresponding unitary block $e^{-i \theta_i \sigma_i}$ in the ans\"atze. This result has been discussed in subsection~\ref{sec:cut-off using Hamiltonian's symmetry}. 

The similar procedure can be applied to discrete symmetries in the statistical models introduced in subsection~\ref{sec:Discrete symmetry}. The Pauli strings reduced by symmetry and the resulting ans\"atze are presented in the next section.

\subsection{Example on continuous symmetry}\label{sec:Example on continuous symmetry}

\begin{table}[!t]
\begin{tabular}{c|c}
U(1) symmetry & U(1) + TR symmetry \\ \hline\hline
XY$-$YX         & \multirow{4}{*}{XY$-$YX}           \\ \cline{1-1}
XX$+$YY         &   \\ \cline{1-1}
ZZ            &                    \\ \cline{1-1}
IZ, ZI        &                    \\ \hline
\end{tabular}
\caption{Table of Pauli strings preserving $U(1)$ symmetry and TR symmetry for one-body interaction term acting on two qubits.}
\label{table:u1-symmetry}
\end{table}

We demonstrate an example of continuous symmetry of the particle number preserving system with Hamiltonian $H_{\mathrm{hop}}$ in Eq.~(\ref{eq:hopping-Hamiltonian}). The system possesses the global $U(1)$ symmetry introduced in subsection~\ref{sec:Continuous symmetry}. The unitary representation of $U(1)$ symmetry with J-W transformation has been given in Eq.~(\ref{eq:U(1)-representation}). In the Hamiltonian $H_{\mathrm{hop}}$, each one-body interaction term $t_{i,i+1}$ acts on two qubits, the unitary representation restricted on the two-qubit Hilbert space reads
\begin{align}
    U_{g_{\alpha}}^{\operatorname{J-W}}=\left( \begin{array}{cccc}
    1 & 0 & 0 &0 \\
    0 & e^{i\alpha} & 0 &0 \\
    0 & 0 & e^{i\alpha} &0 \\
    0 & 0 & 0 &e^{i2\alpha}
        \end{array} \right)\begin{array}{l}
        \ket{00}\\
        \ket{01}\\
        \ket{10}\\
        \ket{11}
        \end{array},
\end{align}
where the corresponding computational basis vectors are shown explicitly in the right column. For each two-qubit Pauli string, one can calculate the commutation relation Eq.~(\ref{eq:Transformation in Pauli-basis}) so that the matrix $c^{(\alpha)}$ can be extracted. Using Eq.~(\ref{eq:master-formula-unitary}), we get constraints on the $4^2=16$ expansion coefficients
\begin{widetext}
\begin{equation}
\begin{aligned}
    a_i&=a_i,  i\in\{II,IZ,ZI,ZZ\};\\
    \left( \begin{array}{cc}
    \cos \alpha & -\sin \alpha \\
    \sin \alpha & \cos \alpha
    \end{array} \right)\left( \begin{array}{c}
    a_i\\
    a_j
    \end{array} \right)&=\left( \begin{array}{c}
    a_i\\
    a_j
    \end{array} \right), (i,j)\in\{(IX,IY),(XI,YI),(ZX,ZY),(XZ,YZ)\};\\
    \left( \begin{array}{cc}
    \cos 2\alpha & -\sin 2\alpha \\
    \sin 2\alpha & \cos 2\alpha
    \end{array} \right)\left( \begin{array}{c}
    a_{XY}+a_{YX}\\
    a_{YY}-a_{XX}
    \end{array} \right)&=\left( \begin{array}{c}
    a_{XY}+a_{YX}\\
    a_{YY}-a_{XX}
    \end{array} \right).
\end{aligned}
\end{equation}
\end{widetext}
As $\alpha$ is an arbitrary real number, these equations lead to 
\begin{equation}
\begin{aligned}
    a_i&=0, \\
    i\in \{IX,IY,XI,YI&,ZX,ZY,XZ,YZ\};\\
    a_{XY}&=-a_{YX};\\
    a_{YY}&=a_{XX},
    \label{eq:U1-solved-result}
\end{aligned}
\end{equation}
with the remaining $a_i, \in\{II,IZ,ZI,ZZ\}$ as free coefficients. We summarize the derived particle number preserving Pauli strings in Table~\ref{table:u1-symmetry}.

Pauli exponentials following the constraints in Eq.~(\ref{eq:U1-solved-result}) can be interpreted intuitively. For example, 
\begin{align}
    e^{i\theta(XX+YY)}=\left( \begin{array}{cccc}
    1 & 0 & 0 &0 \\
    0 & \cos 2\theta & i\sin 2\theta &0 \\
    0 & i\sin 2\theta &\cos 2\theta &0 \\
    0 & 0 & 0 & 1
        \end{array} \right),\\
        e^{i\theta(YX-XY)}=\left( \begin{array}{cccc}
    1 & 0 & 0 &0 \\
    0 & \cos 2\theta & \sin 2\theta &0 \\
    0 & -\sin 2\theta &\cos 2\theta &0 \\
    0 & 0 & 0 & 1
        \end{array} \right),
\end{align}
which are $R_X, R_Y$ rotation within $\{\ket{01},\ket{10}\}$ subspace, respectively.

In the quantum chemistry models with Hamiltonian $H_\mathrm{chem}$ in Eq.~(\ref{eq:quantum-chemistry-model}), after J-W transformation, the two-body interaction term $\hat{a}_i^{\dagger}\hat{a}_j^{\dagger}\hat{a}_k\hat{a}_l$ acts on at least four qubits. To design the ans\"atze for $H_\mathrm{chem}$, we should consider Pauli strings acting on at least four qubits beyond the ones shown in Table~\ref{table:u1-symmetry}. In Appendix~\ref{app:Particle-number-preserving-table}, as an extension of Table~\ref{table:u1-symmetry}, we present a table of particle number preserving Pauli strings on 3 and 4 qubits. Using those Pauli strings, one can design the particle number preserving ansatz carrying out QITE of the quantum chemistry Hamiltonian $H_\mathrm{chem}$.

\subsection{Example on time-reversal symmetry}

By appropriately choosing the basis, we can make the Hamiltonian purely real if the system preserves TR symmetry. So that the local interaction terms of the Hamiltonian commutes with complex conjugation operator
\begin{align}
    [K, H_m]=0.
\end{align}
We further choose a purely real initial state, such as $\ket{\tilde{+}}$ state for the Ising model or the Hartree-Fock state for the quantum chemistry model~\cite{McArdle_20}. So that Eq.~(\ref{eq:Invariance of the initial state}) is satisfied. Finally, by investigating transformation relation Eq.~(\ref{eq:Transformation in Pauli-basis}) for $K$ operator, One can separate Pauli strings into two sets: Pauli strings with an even number of Pauli-$Y$ letters, $P_{even}$, and Pauli strings with an odd number of Pauli-$Y$ letters, $P_{odd}$. Entries of $c^{(\mathrm{TR})}$ matrix are different for these two sets
\begin{align}
    c^{(\mathrm{TR})}_{ij}=\left\{\begin{array}{rl}
        \delta_{i,j}, & \sigma_i \in P_{even};\\
        -\delta_{i,j}, & \sigma_i \in P_{odd}.
    \end{array}\right.
    \label{eq:cm-for-time-reversal-symmetry}
\end{align}
Combining this matrix with Eq.~(\ref{eq:master-formula}), where $\xi=-1$ for the anti-unitary operator, we conclude that only Pauli strings with an odd number of Pauli-$Y$ letters should be involved in the ans\"atze. This is consistent with intuitions because we use unitary transformation $e^{-i\theta\sigma}$ to simulate imaginary time propagator $e^{-\tau H}$. The imaginary time propagator is purely real if all the entries in the Hamiltonian are real. So we need an odd number of Pauli-$Y$ letters to make the Pauli string $\sigma$ purely imaginary.

Ans\"atze carrying out imaginary time evolution for systems with additional TR symmetry can be further simplified. The example for the Ising system possessing both $\mathcal{Z}_2$ and TR symmetries has been given in subsection~\ref{sec:ansatz construction and evolution}. The ans\"atze for the one-body interaction term $t_{i,i+1}$ can be further simplified, as shown in Table~\ref{table:u1-symmetry}.

\begin{figure}
    \centering
    \includegraphics[width=0.48\textwidth]{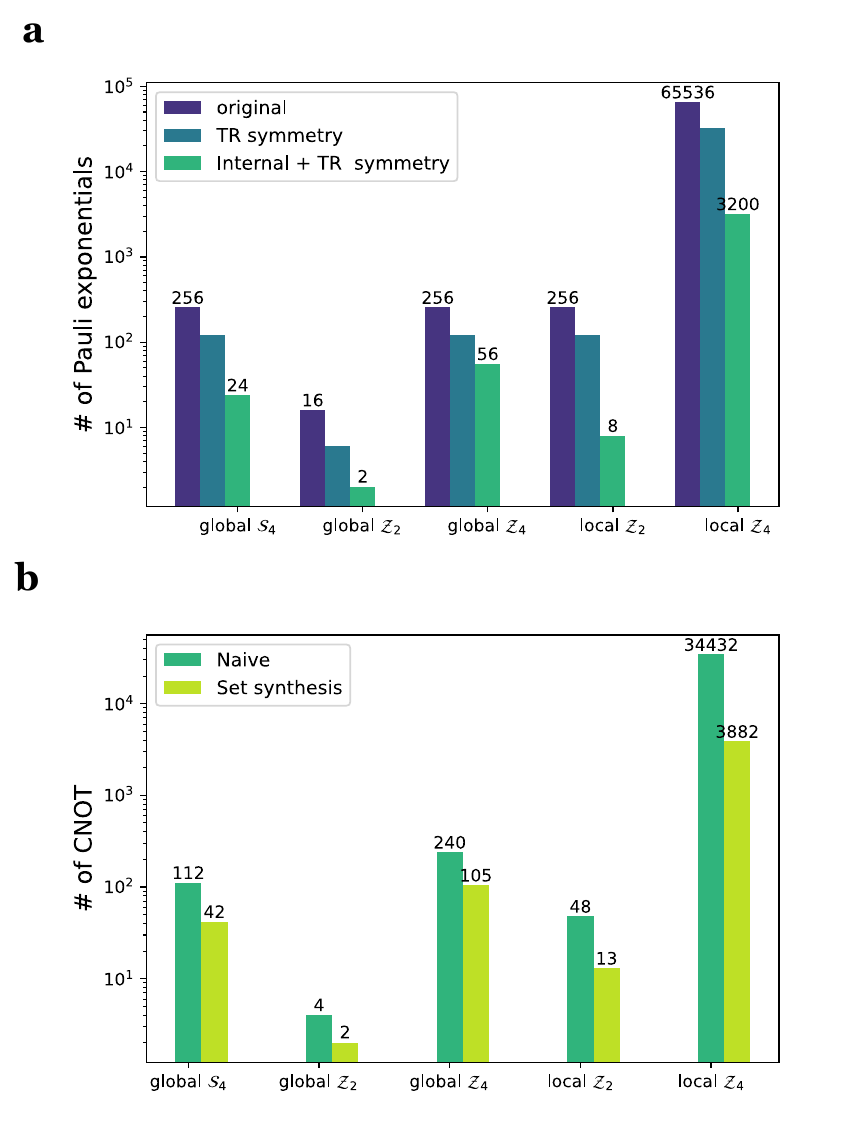}
    \caption{(\textbf{a}) The number of relevant Pauli exponentials reduced by TR and Internal + TR symmetry for different statistical models. The model is labelled by their corresponding internal symmetry in the x-axis, i.e., global $\mathcal{S}_4$: 4-state Potts model, global $\mathcal{Z}_2$: Ising model, global $\mathcal{Z}_4$: 4-state clock model, local $\mathcal{Z}_2$: $\mathcal{Z}_2$ gauge model, local $\mathcal{Z}_4$: $\mathcal{Z}_4$ gauge model. The y-axis is in the log scale.(\textbf{b}) Comparison of the number of CNOT gates by compiling the relevant Pauli exponentials using two compilation strategies: Naive decomposition and Set synthesis. The complied Pauli exponentials have been reduced by Internal + TR symmetry. The y-axis is in the log scale.}
    \label{fig:relevant-paulis}
\end{figure}

\begin{figure*}
    \centering
    \includegraphics[width=0.98\textwidth]{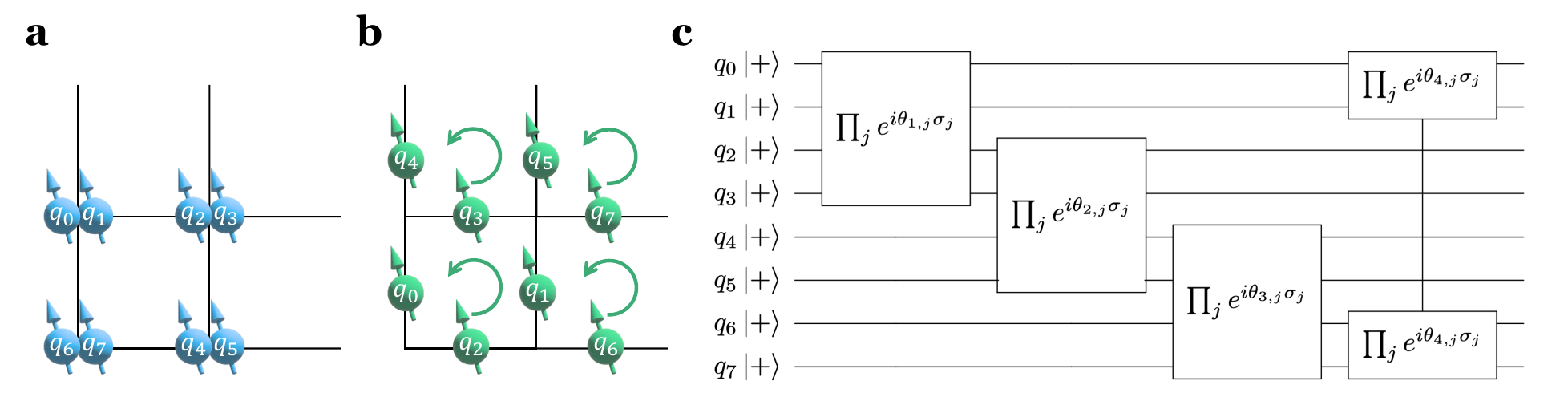}
     \caption{Illustration of qubits layout for the (\textbf{a}) 4-state Potts model, (\textbf{a}) 4-state clock model and (\textbf{b}) $\mathcal{Z}_2$ gauge model on a $2\times 2$ lattice. (\textbf{c}) The symmetry preserving ans\"atze to prepare Gibbs state of the three statistical models. The correspondence of qubit $q_0,\ldots,q_7$ in the statistical model and the quantum circuit is shown explicitly. The circuit shows one layer of the ansatz. Each layer has four blocks corresponding to the four local interaction terms of the model. Each block consists of relevant Pauli exponentials, which respect the symmetries of the corresponding model.}
    \label{fig:DetQITE-inspired-ansatz}
\end{figure*}

\section{Numerical results}\label{sec:Numerical simulations}

In this section, we carry out the VarQITE algorithm on the ans\"atze preserving symmetries for the Potts, clock, and $\mathcal{Z}_Q$ gauge model. The Hamiltonians of these models are introduced in subsection~\ref{sec:Discrete symmetry}. We choose $\ket{\tilde{+}}$ state defined in Eq.~(\ref{eq:Ising-initial-state}) as the initial state. This initial state preserves all the permutation transformation between computational basis. Thus, it is invariant to the symmetry transformation in global $\mathcal{S}_Q$, global $\mathcal{Z}_Q$ and local $\mathcal{Z}_Q$ group and the requirement of Eq.~(\ref{eq:Invariance of the initial state}) is satisfied. Additionally, the thermal expectation values of these statistical models can be derived directly using the imaginary time evolution of $\ket{\tilde{+}}$

\begin{align}
    \ket{\tilde{+}(\tau)}=\frac{e^{-\tau H}\ket{\tilde{+}}}{||e^{-\tau H}\ket{\tilde{+}}||},
\end{align}
as explained in the following.

\begin{table*}[!t]
\begin{tabular}{l|l|c|c}

Model name          & symmetry  & local interaction term& relevant Pauli strings                                                                                                                                                                     \\ \hline\hline
Ising model         & global $\mathcal{Z}_2$ & ZZ                     & ZY,YZ                                                                                                                                                                               \\ \hline
4-state Potts model & global $\mathcal{S}_4$ & IIII+ZIZI+IZIZ+ZZZZ    & \begin{tabular}[c]{@{}c@{}}IYIZ, IZIY, YIZI, ZIYI,\\ IYXZ, IZXY, XYIZ, XZIY, YIZX, YXZI, ZIYX, ZXYI,\\ XYXZ, XZXY, YXZX, ZXYX, \\ ZZZY, ZZYZ, ZYZZ, YZZZ,\\ YYYZ, YYZY, YZYY, ZYYY\end{tabular} \\ \hline
$\mathcal{Z}_2$ gauge model     & local $\mathcal{Z}_2$  & ZZZZ                   & \begin{tabular}[c]{@{}c@{}}ZZZY, ZZYZ, ZYZZ, YZZZ,\\ YYYZ,YYZY,YZYY,ZYYY\end{tabular}                                                                                               \\ \hline
\end{tabular}
\caption{The internal symmetry, local interaction term and relevant Pauli strings reduced by Internal+TR symmetry for Ising model, 4-state Potts model and $\mathcal{Z}_2$ gauge model. The order of Pauli letters in each relevant Pauli string follows that in the local interaction term.}
\label{tab:relevant-Paulis}
\end{table*}

The thermal expectation value for an observable $O$ is defined by
\begin{align}
    \langle O\rangle_{\beta} \equiv \frac{\Tr(Oe^{-\beta H})}{\mathbf{Z}},\mathbf{Z}=\Tr(e^{-\beta H}),
\end{align}
where $\mathbf{Z}$ is the partition function of the thermal system and $\beta$ is the inverse temperature. In Ref.~\cite{Wang23}, the authors show that the thermal expectation value can be calculated with the expectation value of $\ket{\tilde{+}(\tau)}$ state

\begin{align}
    \langle O\rangle_{\beta}=\bra{\tilde{+}(\tau)}O\ket{\tilde{+}(\tau)},
\end{align}
with the inverse temperature $\beta=2\tau$.

Table~\ref{tab:relevant-Paulis} presents Pauli exponentials satisfying the symmetry constraints of the statistical models for the corresponding local interaction terms. In the ans\"atze, we discard Pauli exponentials whose expansion coefficients $a_i=0$ and assign each remaining Pauli exponential a free variational parameter. We call the remaining Paulis exponentials (strings) as the \textit{relevant Pauli exponentials (strings)}. The number of relevant Pauli exponentials characterizes the depth and degree of freedom of the ans\"atze. In Figure~\ref{fig:relevant-paulis}\textbf{a}, we compare the number of relevant Pauli exponentials for one local interaction term of different models before and after the symmetry reductions. The models we considered possess global $\mathcal{S}_4$, global $\mathcal{Z}_2$, global $\mathcal{Z}_4$, local $\mathcal{Z}_2$ and local $\mathcal{Z}_4$ symmetry (We call these symmetries \textit{internal symmetries}, in contrast to TR symmetry), which correspond to 4-state Potts model, Ising model, 4-state clock model, $\mathcal{Z}_2$ gauge model and $\mathcal{Z}_4$ gauge model respectively. The dark blue bars with legend \textit{original} denote Pauli exponentials that have the same length as the local interaction term $|H_m|$, such that
\begin{align}
    \mathrm{original:}\quad \# \mathrm{~of~Pauli~exponentials}=4^{|H_m|}.
\end{align}
The other two colours of bars present the number of relevant Pauli exponentials reduced according to the TR and Internal+TR symmetry. After the symmetry reductions, we see that the number of relevant Pauli exponentials is significantly reduced for different statistical models.

\begin{figure*}[!t]
     \centering
     \begin{subfigure}{0.33\textwidth}
         \centering
         \includegraphics[width=\textwidth]{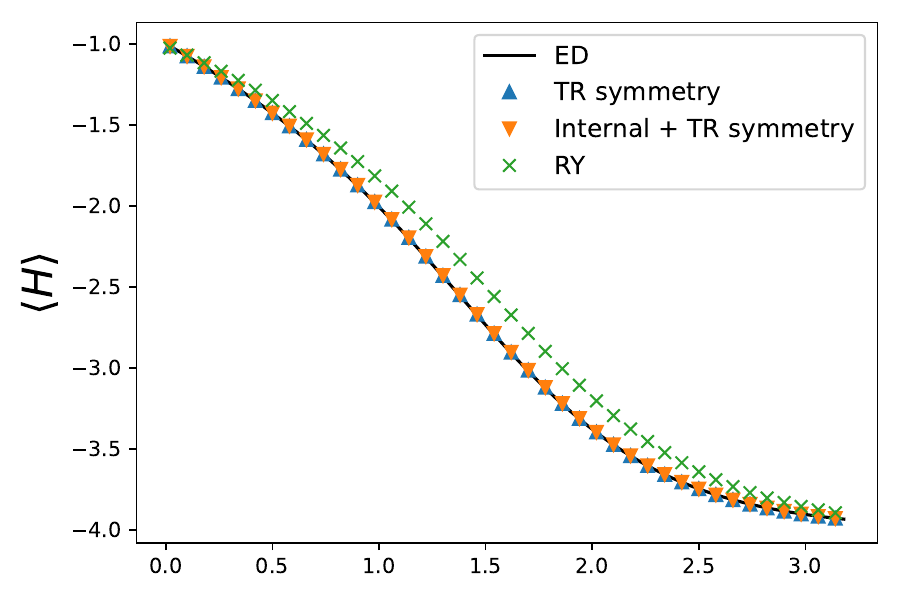}
         \includegraphics[width=\textwidth]{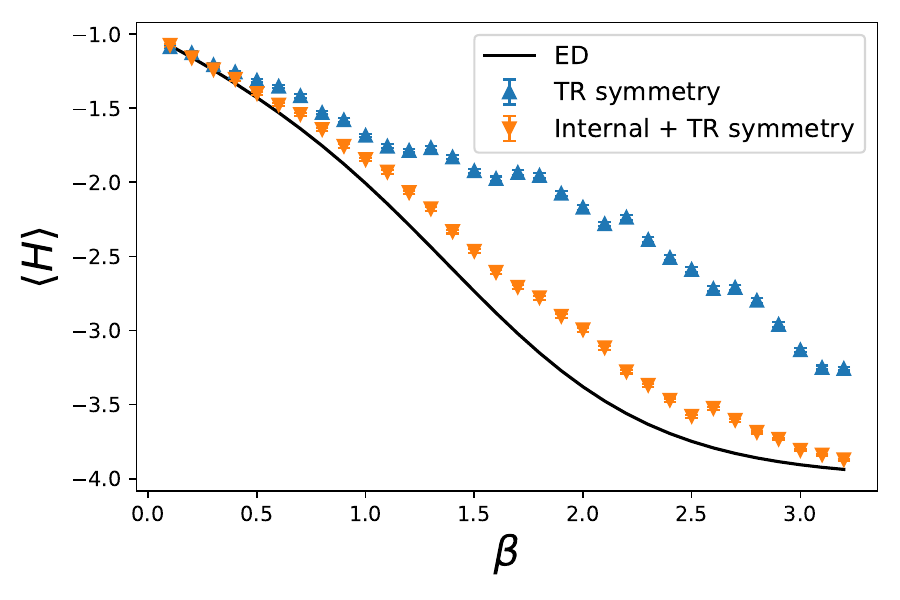}
         \caption{4-state Potts model}
         \label{fig:Potts}
     \end{subfigure}%
     \begin{subfigure}{0.33\textwidth}
         \centering
         \includegraphics[width=\textwidth]{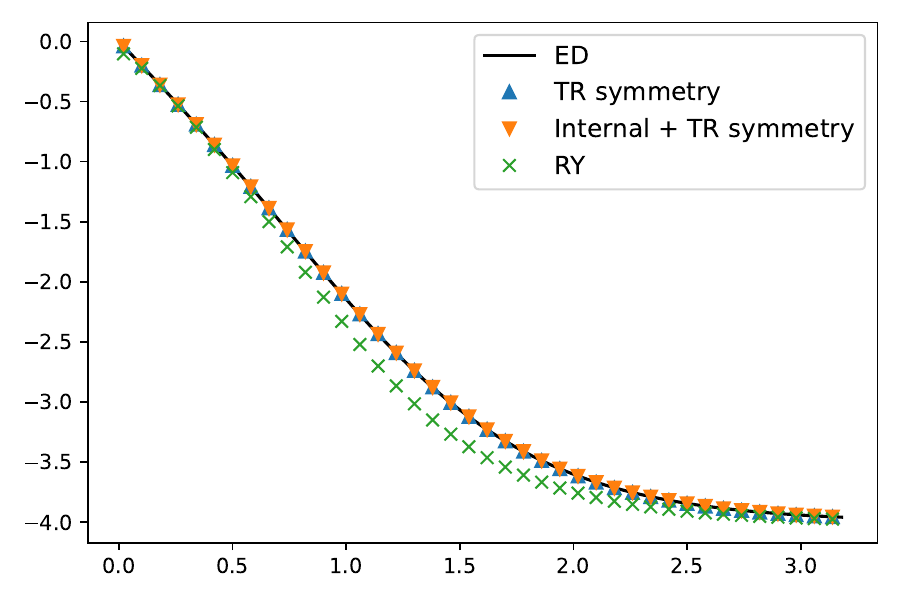}
         \includegraphics[width=\textwidth]{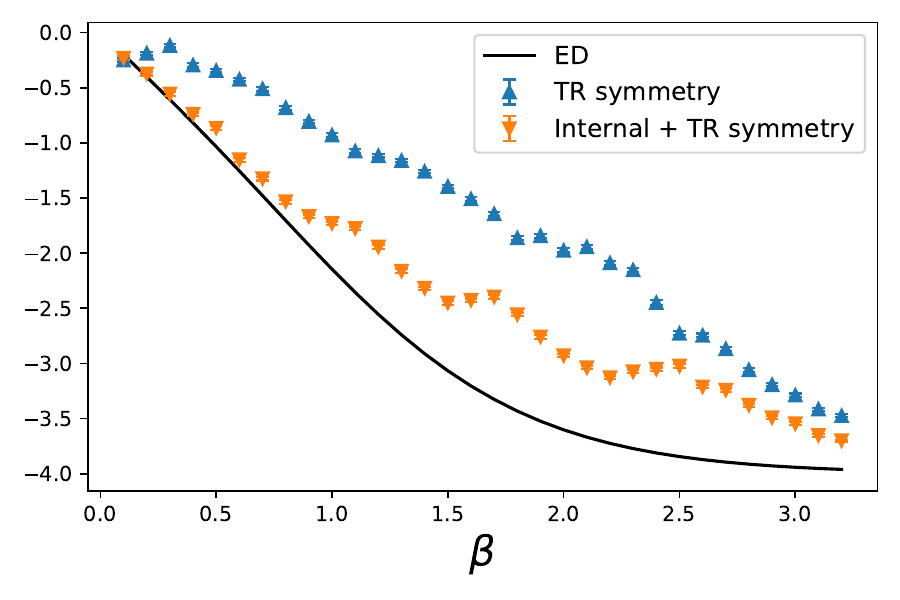}
         \caption{4-state clock model}
         \label{fig:clock}
     \end{subfigure}%
     \begin{subfigure}{0.33\textwidth}
         \centering
         \includegraphics[width=\textwidth]{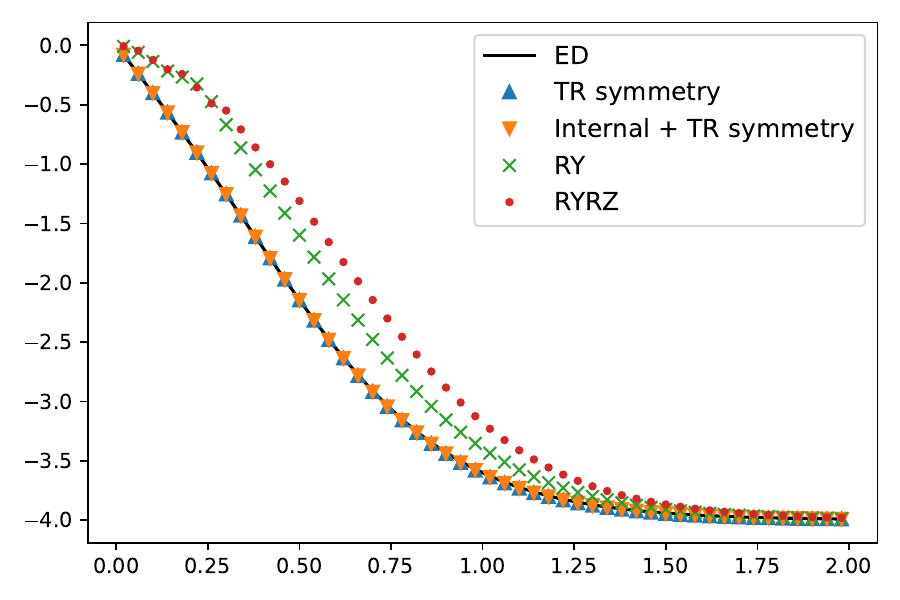}
         \includegraphics[width=\textwidth]{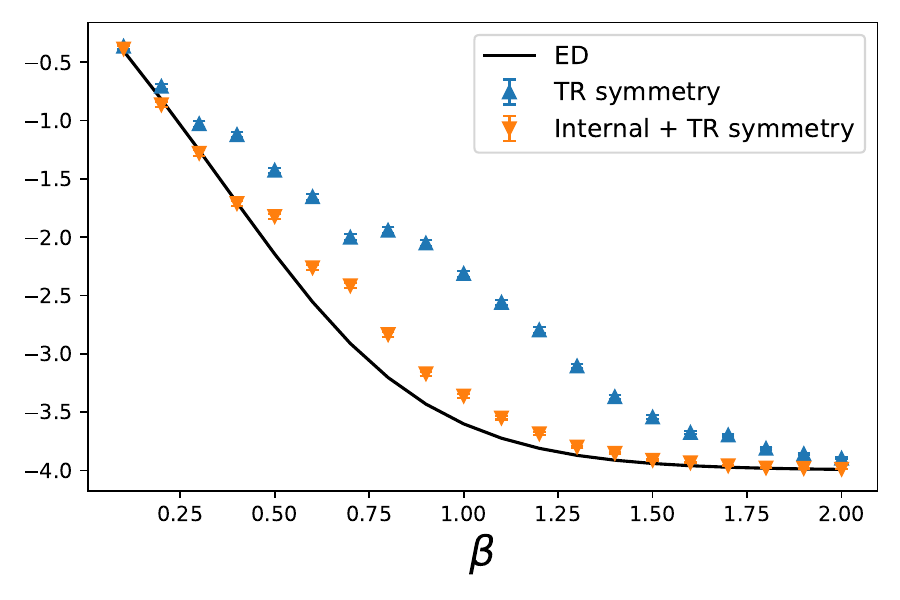}
         \caption{$\mathcal{Z}_2$ gauge model}
         \label{fig:gauge}
     \end{subfigure}
        \caption{Numerical results of estimated thermal energy using VarQITE algorithm, for the (\textbf{a}) 4-state Potts model, (\textbf{b}) 4-state clock model and (\textbf{c}) $\mathcal{Z}_2$ gauge model. The upper (lower) row shows the simulation results without(with) shot noise. In the upper row, we compare the results obtained from our proposed ans\"atze and the hardware efficient ans\"atze RY and RYRZ. The proposed ans\"atze are reduced by implementing TR symmetry and Internal+TR symmetry, and have one layer $L=1$. Their corresponding results are denoted by upper triangle and lower triangle, respectively. Results using RY and RYRZ ans\"atze are denoted by cross and circle, respectively. ED denotes results from exact diagonalization. In the lower row, we see that the results from the ans\"atze constrained by Internal+TR symmetry are more consistent with the ED results, compared to the ones reduced by only TR symmetry. In the VarQITE algorithm, we adopt the Euler step length $\delta \tau=0.01$ ($0.05$) for the simulation in the upper (lower) row.}
        \label{fig:energy-expectation}
\end{figure*}

\begin{table*}[!t]
\begin{tabular}{l|l|l|l}

Model name                     & Ans\"atze & \# CNOT & \# parameters \\ \hline\hline
\multirow{3}{*}{4-state Potts model (Figure~\ref{fig:Potts})} & TR symmetry              &   1984  & 480           \\
                               & Internal+TR symmetry     & 168     & 96            \\ 
                                & RY                       & 168     & 200           \\
                               \hline
\multirow{3}{*}{4-state clock model (Figure~\ref{fig:clock})} 
                                & TR symmetry           &   1984  & 480           \\
                               & Internal+TR symmetry     & 420     & 224           \\ 
                               & RY                       & 420     & 488           \\ \hline
\multirow{4}{*}{$\mathcal{Z}_2$ gauge model (Figure~\ref{fig:gauge})}     
                               & TR symmetry              &  1984   & 480           \\ 
                               & Internal+TR symmetry     & 52      & 32            \\
                               & RY                       & 56      & 72            \\
                                & RYRZ                     & 56      & 144           \\ \hline
\end{tabular}
\caption{Number of CNOT gates and parameters of the ans\"atze used in the simulations of Figure~\ref{fig:energy-expectation}.}
\label{table:cnot-parameters-counts}
\end{table*}

To realize the relevant Pauli exponentials on the quantum circuit, we need to compile the Pauli exponentials using single-qubit rotation gates and two-qubit CNOT gates. In Figure~\ref{fig:relevant-paulis}\textbf{b}, we present the number of CNOT gates required to realize the relevant Pauli exponentials after Internal+TR symmetry reduction. We compare two compilation strategies: Naive decomposition~\cite{Nielsen2000} and Set synthesis proposed in Ref.~\cite{Cowtan2020AGC}. Using Set synthesis, the number of CNOT gates for the relevant Pauli exponentials can be vastly reduced, especially when the number of CNOT gates is plenty in the Naive decomposition, such as the case for the $\mathcal{Z}_4$ gauge model. We will use these CNOT gate counts to compare the symmetry-preserving ans\"atze with other frequently-used ans\"atze.

\begin{figure}[h]
    \centering
    \includegraphics[width=0.48\textwidth]{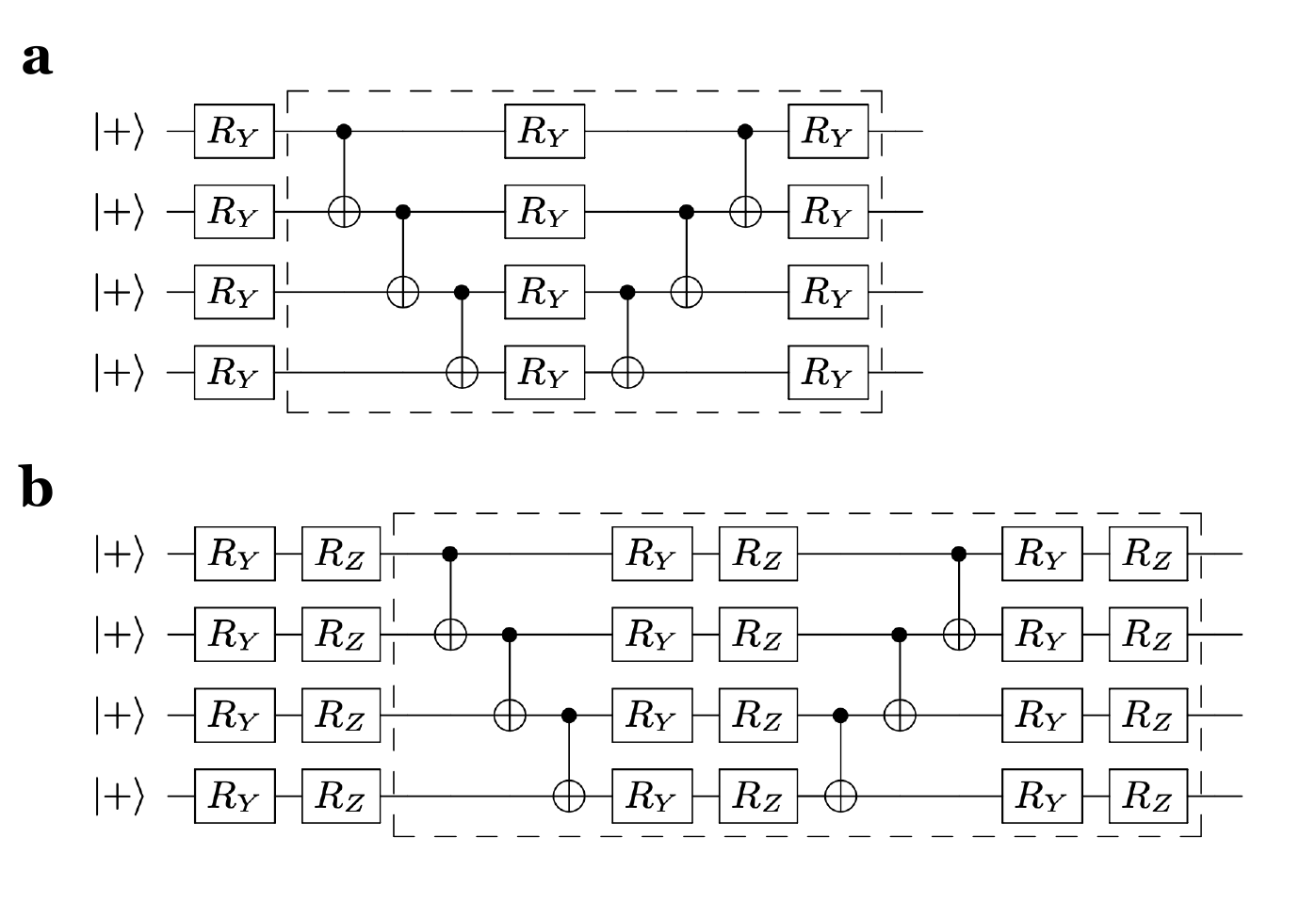}
     \caption{Illustration of the hardware-efficient ans\"atze (\textbf{a}) RY and (\textbf{b}) RYRZ used in the simulations of Fi-gure~\ref{fig:energy-expectation}. The dashed box in the circuit denotes one la-yer of the ansatz. Here we demonstrate the case of 4 qubits. We use the circuits with eight qubits following similar patterns in the numerical simulations.}
    \label{fig:hardware-efficient-ansatz}
\end{figure}

We carry out the VarQITE algorithm for the 4-state Potts model, 4-state clock model and $\mathcal{Z}_2$ gauge model using the proposed ans\"atze on the statevector quantum simulator~\cite{Qiskit}. The ans\"atze are illustrated in Figure~\ref{fig:DetQITE-inspired-ansatz}, which simulates a lattice of size $2\times 2$, with $4$ nodes and $8$ links. We assign $2$ qubits for each node in the 4-state Potts model and 4-state clock model, and one qubit for each link in the $\mathcal{Z}_2$ gauge model. So all the simulations have $8$ qubits. The ans\"atze have layer $L=1$ in all the simulations. The resulting energy expectation values are shown in the upper row of Figure~\ref{fig:energy-expectation}. We compare the results obtained from the frequently-used hardware efficient ans\"atze (RY, RYRZ). These two ans\"atze are illustrated in Figure~\ref{fig:hardware-efficient-ansatz} (for 4-qubit case). The dashed box in each circuit denotes one layer of the ans\"atze. We tune the number of layers in the hardware efficient ans\"atze so that their numbers of CNOT gates and free parameters surpass those of our proposed ans\"atze. CNOT gates and free parameter counts are shown explicitly in Table~\ref{table:cnot-parameters-counts}. It is clear that for all the statistical models, the thermal expectation energy obtained using our proposed ans\"atze is more consistent with exact diagonalization (ED) results, while our proposed ans\"atze use less number of CNOT gates and free parameters.

In the upper row of Figure~\ref{fig:energy-expectation}, differences between the results before and after implementing the constraints of internal symmetries are not obvious. It means that the ans\"atze have the same ability to carry out the imaginary time evolution before and after the symmetry reduction. Meanwhile, the internal-symmetry-reduced ans\"atze have fewer free parameters, which is beneficial to the evolution process under noise. To see this, in the lower row of Figure~\ref{fig:energy-expectation}, we consider shot noise in the calculation of $\widetilde{M}$ matrix and $\widetilde{V}$ vector, i.e., we add a Gaussian random number in each entry of $\widetilde{M}$ and $\widetilde{V}$. The shot noise introduces instability in every evolution step of VarQITE, which is inevitable in real quantum computation. The evolution is more unstable with more free parameters in the ansatz, since $\widetilde{M}$ and $\widetilde{V}$ would have larger dimensions. The lower row of Figure~\ref{fig:energy-expectation} shows that, after reducing the number of free parameters according to the internal symmetries, the thermal expectation energies are more consistent with the exact diagonalization (ED) results.

\section{Conclusion and outlook}
We propose a systematic method of implementing symmetries in constructing ans\"atze for the VarQITE algorithm. This method can be applied to both unitary and anti-unitary symmetries in quantum mechanical systems. We show some examples of ans\"atze preserving discrete or continuous symmetries that can be designed using our method. Using numerical simulations, we demonstrate that the symmetry-enhanced ans\"atze have good expressivity and outperform the frequently-used hardware-efficient ans\"atze in Gibbs state preparation. We show that, after the symmetry reduction, the ans\"atze are more robust to the shot noise during the imaginary time evolution.

In this work, we prepare the Gibbs state using the proposed ans\"atze and VarQITE algorithm. For preparation of the ground state on quantum computers, one can also use variational quantum eigensolver (VQE)~\cite{Peruzzo:2014, McClean_17} algorithm. The ans\"atze proposed in this work can also be applied in VQE algorithm, which could perform better than the frequently-used Hamiltonian variational ansatz. Because the Hamiltonian variational ans\"atze are designed to carry out the real time evolution of adiabatic processes, while our ans\"atze carry out the imaginary one. The imaginary time evolution could converge faster to the ground state than the real time evolution of adiabatic processes~\cite{Farhi2000QuantumCB}.

More complicated symmetry groups and physical models can be explored utilizing the symmetry-preserving method. For example, the point groups in material science and the non-Abelian Lie groups in high-energy physics can be considered. In the main text, we apply the method to design ans\"atze for particle number preserving systems with one- and two-body interaction terms and Jordan-Wigner encoding. One can consider quantum chemical systems with many-body interactions and other fermionic encoding methods, such as Bravyi-Kitaev and Verstraete-Cirac encoding~\cite{Bravyi_02, Verstraete:2005pn}. We leave those considerations for future works.

\section{Acknowledgements}
X.W. and X.F. were supported in part by NSFC of China under Grants No. 12125501, No. 12070131001, and No. 12141501, and National Key Research and Development Program of China under No. 2020YFA0406400. C.T.\ is supported in part by the Helmholtz Association - ``Innopool Project Variational Quantum Computer Simulations (VQCS)''. This work is supported with funds from the Ministry of Science, Research and Culture of the State of Brandenburg within the Centre for Quantum Technologies and Applications (CQTA). This work is funded by the European Union’s Horizon Europe Framework Programme (HORIZON) under the ERA Chair scheme with grant agreement No.\ 101087126.\\

\begin{center}
    \includegraphics[width = 0.08\textwidth]{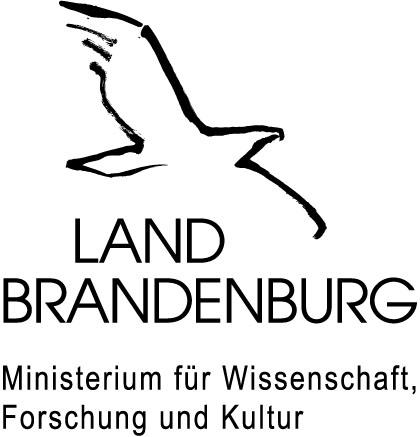}
\end{center}

\appendix

\section{Statistical models encoding on quantum computers}\label{app:Statistical models and their representation}

In this Appendix, we show the encoding of the statistical models on quantum computers and their internal symmetries. The representations of the symmetry operations on the computational basis are utilized to reduce the number of Pauli exponentials. Here, we discuss the $Q$-state Potts model, $Q$-state clock model and $\mathcal{Z}_Q$ gauge model, as introduced in the main text, which has the global $\mathcal{S}_Q$, global $\mathcal{Z}_Q$ and local $\mathcal{Z}_Q$ symmetries, respectively.

\subsection{Potts model}
The Hamiltonian of the $Q$-state Potts model~\cite{Wu_82} reads
\begin{align}
    H_{\mathrm{Potts}}=-\sum_{\langle n,n'\rangle}\delta(k_{n},k_{n'}),
    \label{eq:potts-hamiltonian-app}
\end{align}
where $k_{n},k_{n'} \in \{0,1,\ldots, Q-1\}$ is a label of states. $n,n'$ denote nodes on a square lattice and $\langle n,n'\rangle$ denotes nearest neighbor coupling. We use binary encoding to encode the $Q$ states on the computational basis, i.e., 
\begin{equation}
\begin{aligned}
    0&\rightarrow \ket{00\ldots 00},\\
    1&\rightarrow \ket{00\ldots 01},\\
    &\vdots\\
    Q-1&\rightarrow \ket{11\ldots 11}.
    \label{eq:binary-encoding}
\end{aligned}
\end{equation}
For convenience, we only consider the $Q=2^P$-state Potts model, where $P$ is an integer. By interpreting $\delta(k_{n},k_{n'})$ as a projector in the computational basis, the Potts Hamiltonian can be rewritten using Pauli-$Z$ strings
\begin{align}
     H_{\mathrm{Potts}} = -\sum_{\langle n,n'\rangle}\prod_{p=0}^{P-1}\left(\frac{1+Z_{n,p}Z_{n',p}}{2}\right).
\end{align}
The local interaction term of this Hamiltonian is
\begin{align}
    H_{n,n'}= \prod_{p=0}^{P-1}\left(\frac{1+Z_{n,p}Z_{n',p}}{2}\right).
\end{align}

The symmetry operations are permutations between all the state labels among all the lattice sites, constituting the global permutation group $\mathcal{S}_Q$. Consider the $4=2^2$-state Potts model as an example. The $\mathcal{S}_4$ group has generators
\begin{align}
    \mathcal{S}_4=\langle(0,1),(1,2),(2,3),(3,0)\rangle,
\end{align}
which are adjacent transpositions. On a single lattice site $n$, they can be represented using a unitary matrix under the computational basis
\begin{align}
    U_{(0,1),n}=\left( \begin{array}{cccc}
    0 & 1 & 0 &0 \\
    1 & 0 & 0 &0 \\
    0 & 0 & 1 &0 \\
    0 & 0 & 0 &1
        \end{array} \right)\begin{array}{c}
        \ket{00}\\
        \ket{01}\\
        \ket{10}\\
        \ket{11}
        \end{array},
\end{align}
\begin{align}
    U_{(1,2),n}=\left( \begin{array}{cccc}
    1 & 0 & 0 &0 \\
    0 & 0 & 1 &0 \\
    0 & 1 & 0 &0 \\
    0 & 0 & 0 &1
        \end{array} \right)\begin{array}{c}
        \ket{00}\\
        \ket{01}\\
        \ket{10}\\
        \ket{11}
        \end{array},
\end{align}
\begin{align}
    U_{(2,3),n}=\left( \begin{array}{cccc}
    1 & 0 & 0 &0 \\
    0 & 1 & 0 &0 \\
    0 & 0 & 0 &1 \\
    0 & 0 & 1 &0
        \end{array} \right)\begin{array}{c}
        \ket{00}\\
        \ket{01}\\
        \ket{10}\\
        \ket{11}
        \end{array},
\end{align}
\begin{align}
    U_{(3,0),n}=\left( \begin{array}{cccc}
    0 & 0 & 0 &1 \\
    0 & 1 & 0 &0 \\
    0 & 0 & 1 &0 \\
    1 & 0 & 0 &0
        \end{array} \right)\begin{array}{c}
        \ket{00}\\
        \ket{01}\\
        \ket{10}\\
        \ket{11}
        \end{array},
\end{align}
where the corresponding basis vectors are shown explicitly. Then the global symmetry operation can be represented as
\begin{align}
    U_g= \prod_{n}\; U_{g,n}, \forall g\in \mathcal{S}_4,
\end{align}
which is similar to the symmetry operation Eq.~(\ref{eq:Ising-symmetry-operation}) of Ising model. Ising model is a special case of the $Q=2$-state Potts model.
        
\subsection{Clock model}
The Hamiltonian of clock model~\cite{Wu_82} reads 
\begin{align}
    H_{\mathrm{clock}} = -\sum_{\langle n,n'\rangle}\cos (\theta_{n}-\theta_{n'}),
    \label{eq:clock-model-app}
\end{align}
where $\theta_{n}\equiv 2\pi k_{n}/Q, k_{n} \in \{0,1,\ldots, Q-1\}$, and also $Q=2^P$. This Hamiltonian can be encoded on the computational basis as follows. The clock Hamiltonian can be rewritten into a real part of exponentials
\begin{align}
    H_{\mathrm{clock}}=-\sum_{\langle n,n'\rangle }\Re{e^{i\theta_{n}}\ket{\theta_{n}}\bra{\theta_{n}} e^{-i\theta_{n'}} \ket{\theta_{n'}}\bra{\theta_{n'}} },
    \label{eq:quantum-clock-model}
\end{align}
with the corresponding projectors added explicitly. In our simulation, we use the binary encoding of $\ket{\theta_{n}}$
\begin{align}
    \ket{\theta_{n}}=\ket{2\pi k_{n}/Q}\rightarrow \ket{k_{n (2)}}, k_{n}=0,1,\ldots, Q-1,
\end{align}
where $k_{n (2)}$ represents binary representation of the integer $k_{n}$ like the one shown in Eq.~(\ref{eq:binary-encoding}). With binary encoding, operator $e^{i\theta_{n}}\ket{\theta_{n}}\bra{\theta_{n}}$ in Eq.~(\ref{eq:quantum-clock-model}) can be written more concisely by introducing diagonal operators
\begin{equation}
\begin{aligned}
    R(p)\equiv& \ket{0}\bra{0}+e^{2\pi i/2^p}\ket{1}\bra{1}\\
    =&\frac{1+e^{2\pi i/2^p}}{2}I+\frac{1-e^{2\pi i/2^p}}{2}Z.
\end{aligned}
\end{equation}
Then one finds 
\begin{align}
e^{i\theta_{n}}\ket{\theta_{n}}\bra{\theta_{n}}=\prod_{p=1}^P \otimes R(p).
\end{align}
With these observations, we can write the clock Hamiltonian in the computational basis using Pauli-$Z$ strings.

The symmetry group of $Q$-state clock model is additive group $\mathcal{Z}_Q$, which is a subgroup of $\mathcal{S}_Q$. Consider an example of $4$-state clock model. the $\mathcal{Z}_4$ group has one generator 
\begin{align}
    \mathcal{Z}_4=\langle (0,1,2,3)\rangle.
\end{align}
This generator on a single site $n$ can be represented in the computational basis
\begin{align}
    U_{(0,1,2,3),n}=\left( \begin{array}{cccc}
    0 & 0 & 0 &1 \\
    1 & 0 & 0 &0 \\
    0 & 1 & 0 &0 \\
    0 & 0 & 1 &0
        \end{array} \right)\begin{array}{c}
        \ket{00}\\
        \ket{01}\\
        \ket{10}\\
        \ket{11}
        \end{array}.
        \label{eq:4-state-clock-unitary-operation}
\end{align}
The global symmetry operation is the tensor product of $U_{(0,1,2,3),n}$ on all the spatial nodes.

Techniques introduced here can be easily extended to $\mathcal{Z}_Q$ gauge models. The Hamiltonian $H_{\mathrm{gauge}}$ in Eq.~(\ref{eq:z-q-gauge-model-hamiltonian}) can be written similarly to that in Eq.~(\ref{eq:quantum-clock-model}), and the symmetry operations in local $\mathcal{Z}_Q$ group Eq.~(\ref{eq:z-q-local}) can be represented under computational basis similar to the ones in Eq.~(\ref{eq:4-state-clock-unitary-operation}).

\section{$c^{(\bos{\alpha})}$ matrix}\label{app:cm-matrix}

In this appendix, we will show some basic properties of $c^{(\bos{\alpha})}$ matrix according to its definition
\begin{align}
    U_{g_{\bos{\alpha}}}^{-1} \sigma_i U_{g_{\bos{\alpha}}}=\sum_{j} c^{(\bos{\alpha})}_{ij} \sigma_j.
    \label{eq:U-commutation-relation}
\end{align}
As the $c^{(\bos{\alpha})}$ matrix for the anti-unitary TR operator has been derived explicitly in the main text, here we only consider unitary matrix $U_{g_{\bos{\alpha}}}^{-1}=U_{g_{\bos{\alpha}}}^{\dagger}$.  The properties of $c^{(\bos{\alpha})}$ matrix is listed as follows
\begin{itemize}
    \item[1.] $c^{(\bos{\alpha})}$ is real, i.e., $c^{(\bos{\alpha})}_{ij}=c^{(\bos{\alpha})*}_{ij}$.
    \item[2.] $ U_{g_{\bos{\alpha}}} \sigma_i U_{g_{\bos{\alpha}}}^{-1}=\sum_{i}[c^{(\bos{\alpha})-1}]_{ij}\sigma_{j}$.
    \item[3.] $c^{(\bos{\alpha})}$ is orthogonal, i.e., $c^{(\bos{\alpha})T}=c^{(\bos{\alpha})-1}$.
\end{itemize}
\textit{proof}: The first property can be proved by taking Hermitian conjugate on both sides of Eq.~(\ref{eq:U-commutation-relation}). Then we see $c^{(\bos{\alpha})}_{ij}=c^{(\bos{\alpha})*}_{ij}$ according to the Hermicity of Pauli strings and the orthonormal relation of the Pauli-basis
\begin{align}
    \frac{1}{d}\tr(\sigma_i\sigma_j)=\delta_{ij},
    \label{eq:orthonormal-relation}
\end{align}
where $d$ is the dimension of Pauli strings. The second property can be proved by writing Eq.~(\ref{eq:U-commutation-relation}) as $\sigma_i=\sum_{j} c_{ij}^{(\bos{\alpha})} U_{g_{\bos{\alpha}}} \sigma_{j} U_{g_{\bos{\alpha}}}^{-1}$ and assume $U_{g_{\bos{\alpha}}} \sigma_{j} U_{g_{\bos{\alpha}}}^{-1}=\sum_{k} d_{jk}^{(\bos{\alpha})} \sigma_{k}$, then we find $\sum_j c^{(\bos{\alpha})}_{ij} d_{jk}^{(\bos{\alpha})}=\delta_{ik}$, so that $d^{(\bos{\alpha})}$ is the inverse of $ c^{(\bos{\alpha})}$ matrix. For the third property, notice that $c^{(\bos{\alpha})}_{ij}$ can be derived by combining the orthonormal relation Eq.~(\ref{eq:orthonormal-relation}) with Eq.~(\ref{eq:U-commutation-relation})
\begin{equation}
\begin{aligned}
    c^{(\bos{\alpha})}_{ij}&=\tr(\sigma_j U_{g_{\bos{\alpha}}}^{-1}\sigma_i U_{g_{\bos{\alpha}}})/d\\
    &=\tr(U_{g_{\bos{\alpha}}} \sigma_j U_{g_{\bos{\alpha}}}^{-1}\sigma_i )/d\\
    &=\sum_{j'} \tr( [c^{(\bos{\alpha})-1}]_{jj'} \sigma_{j'} \sigma_i )/d\\
    &=[c^{(\bos{\alpha})-1}]_{ji}.
\end{aligned}
\end{equation}
In the third line, we use the second property, and in the last line, we again use the orthonormal relation Eq.~(\ref{eq:orthonormal-relation}). Thus, we prove the third property. $\qed$

\section{Constraints on the expansion coefficients $\bos{a}_0$}\label{app:theorem-proof}

The theorem in section~\ref{sec:Implementing symmetry in ansatz} states that the expansion coefficients $\bos{a}_0$ satisfy the constraints
\begin{align}
    c^{(\bos{\alpha})}\bos{a}_0=\xi \bos{a}_0,\quad \forall g_{\bos{\alpha}}\in \GG,
    \label{eq:master-formula-appendix}
\end{align}
with $\xi=+1$ for unitary symmetry transformation and $\xi=-1$ for anti-unitary symmetry transformation. Here we provide the proof.

\textit{proof}. For unitary transformation $g_{\bos{\alpha}}$, it has unitary representation $\mathcal{U}_{g_{\bos{\alpha}}} = U_{g_{\bos{\alpha}}}$, where $U_{g_{\bos{\alpha}}}$ is a unitary matrix. Then the vector $V$ is invariant by the transformation $c^{(\bos{\alpha})}$
\begin{equation}
\begin{aligned}
    V_i &= \Im(\bra{U_{g_{\bos{\alpha}}}\psi}\sigma_i H_m \ket{U_{g_{\bos{\alpha}}}\psi})\\
    &= \Im(\bra{\psi}U_{g_{\bos{\alpha}}}^{-1}\sigma_i H_mU_{g_{\bos{\alpha}}} \ket{\psi})\\
    &=\Im(\bra{\psi}U_{g_{\bos{\alpha}}}^{-1}\sigma_i U_{g_{\bos{\alpha}}} H_m \ket{\psi})\\
    &=\sum_{j} c^{(\bos{\alpha})}_{ij} V_j.
\end{aligned}
\end{equation}
In the first line, the vector is invariant due to Eq.~(\ref{eq:Invariance of the initial state}). In the third line, we use the commutation relation Eq.~(\ref{eq:Commutation relation}) and finally, the real coefficients $c^{(\bos{\alpha})}_{ij}$ appear according to Eq.~(\ref{eq:Transformation in Pauli-basis}). Similarly, the $M$ matrix is invariant by the transformation
\begin{align}
M_{ij}=\sum_{kl}c^{(\bos{\alpha})}_{ik}c^{(\bos{\alpha})}_{jl}M_{kl}.
\label{eq:M-transformation}
\end{align}

For anti-unitary transformation $g_{\bos{\alpha}}$, it is represented by $\mathcal{U}_{g_{\bos{\alpha}}}=U_{g_{\bos{\alpha}}}K$, where $K$ is the complex conjugation operator in Eq.~(\ref{eq:anti-unitary operator}). An additional minus sign would appear for the transformation of the $V$ vector
\begin{equation}
\begin{aligned}
    V_i &= \Im(\bra{U_{g_{\bos{\alpha}}}K\psi}\sigma_i H_m \ket{U_{g_{\bos{\alpha}}}K\psi})\\
    &= \Im(\overline{\bra{\psi}K U_{g_{\bos{\alpha}}}^{-1}\sigma_i H_mU_{g_{\bos{\alpha}}} K\ket{\psi}})\\
    &=\Im(\overline{\bra{\psi}\mathcal{U}_{g_{\bos{\alpha}}}^{-1}\sigma_i \mathcal{U}_{g_{\bos{\alpha}}}  H_m \ket{\psi}})\\
    &=-\sum_{j} c^{(\bos{\alpha})}_{ij} V_j.
\end{aligned}
\end{equation}
In the second line, the complex conjugation is taken according to Eq.~(\ref{eq:K-effect}). The minus sign in the last line appears because the $V$ vector takes the imaginary part of the overlap. Transformation of $M$ matrix by anti-unitary operator still follows Eq.~(\ref{eq:M-transformation}) because $M$ matrix takes the real part of the overlap.

Gathering the formulae derived above and Eq.~(\ref{eq:DetQITE-linear-equation-again}), we can write them in a compact form
\begin{align}
    \left\{\begin{array}{rl}
         c^{(\bos{\alpha})}M c^{(\bos{\alpha}) T}&=M;\\
         \xi c^{(\bos{\alpha})}V&=V;\\
         M \bos{a}_0&=V, 
    \end{array}\right.
    \label{eq:key-formulae}
\end{align}
where $(\cdot)^T$ denotes matrix transpose and $\xi=+1$ for unitary transformation and $\xi=-1$ for anti-unitary transformation. Taking the first two formulae to the last one, we have 
\begin{align}
    M c^{(\bos{\alpha}) T} \bos{a}_0=\xi V,
    \label{eq:mctaxiv}
\end{align}
where $c^{(\bos{\alpha})}$ has been cancelled on the both side. Together with $M \bos{a}_0=V$, if $M$ is full rank, we have
\begin{align}
    c^{(\bos{\alpha}) T} \bos{a}_0=\xi \bos{a}_0.
    \label{eq:cTa=xia}
\end{align}
On the other hand, if $M$ is not full rank, because $\bos{a}_0$ is a special solution from pseudo-inverse, Eq.~(\ref{eq:cTa=xia}) still holds. The rigorous proof is provided in Appendix~\ref{app:cTa=xia}. 

As $c^{(\bos{\alpha})}$ is a real orthogonal matrix, we conclude Eq.~(\ref{eq:master-formula-appendix}) from Eq.~(\ref{eq:cTa=xia})$\qed$

We provide two comments on this theorem. They help to simplify the calculation procedure of solving the constraints. The first comment provides proof of the corollary in section~\ref{sec:Implementing symmetry in ansatz}. We focus on the unitary symmetry transformations, i.e.,
\begin{align}
    c^{(\bos{\alpha})}\bos{a}_0=\bos{a}_0,\quad \forall g_{\bos{\alpha}}\in \GG.
    \label{eq:master-formula-unitary-appendix}
\end{align}

First, Eq.~(\ref{eq:master-formula-unitary-appendix}) indicates that we need to solve the linear equations for all the elements in the symmetry group $\GG$. In practice, we only need to consider the symmetry transformation from the generating set of the group. As the number of elements in the generating set is less than that of the whole group, the number of constraints to be solved is reduced. To see this, assume two generators of the symmetry group $g_{\bos{\alpha}},g_{\bos{\beta}}$ and their multiplication $h = g_{\bos{\alpha}}g_{\bos{\beta}}$. Transformation of $h$ on the Pauli-basis reads
\begin{equation}
\begin{aligned}
    U_{h}^{-1} \sigma_i U_{h}&=U_{g_{\bos{\beta}}}^{-1}U_{g_{\bos{\alpha}}}^{-1}\sigma_i U_{g_{\bos{\alpha}}}U_{g_{\bos{\beta}}}\\
    &=\sum_{jk} c_{ij}^{(\bos{\alpha})}c_{jk}^{(\bos{\beta})}\sigma_k.
\end{aligned}
\end{equation}
Thus, the constrain of $h$ on the expansion coefficients is
\begin{align}
    \sum_{jk} c_{ij}^{(\bos{\alpha})}c_{jk}^{(\bos{\beta})}(\bos{a}_0)_k=(\bos{a}_0)_i.
\end{align}
This equation is satisfied automatically if Eq.~(\ref{eq:master-formula-unitary-appendix}) holds for the generator $g_{\bos{\alpha}}$ and $g_{\bos{\beta}}$. Thus, solving the constraint from $h$ is redundant. This conclusion can be generalized to an arbitrary product of generators. Since each element in the group $\GG$ can be written as a product of generators, the corresponding constraint is automatically satisfied by only considering constraints from the generating set of group $\GG$. Thus we prove the corollary in section~\ref{sec:Implementing symmetry in ansatz}.

Second, Eq.~(\ref{eq:master-formula-unitary-appendix}) can be related to the previous works on symmetry preserving ans\"atze. Combining the original definition of $c^{(\bos{\alpha})}$ in Eq.~(\ref{eq:Transformation in Pauli-basis}), Eq.~(\ref{eq:master-formula-unitary-appendix}) can be rewritten as a commutation relation
\begin{align}
    [U_{g_{\bos{\alpha}}}, \hat{A}_0]=0,\quad \forall g_{\bos{\alpha}}\in \GG,
\end{align}
where $\hat{A}_0\equiv \sum_i (\bos{a}_0)_i\sigma_i$, as defined similar to the $\hat{A}$ in Eq.~(\ref{eq:A-definition}). This relation can be directly derived by symmetry constraints to the unitary evolution of a quantum system, which has been discussed in previous works, such as the framework of GQML~\cite{Larocca_22, Meyer_22, Sauvage_22, Nguyen_22, Ragone_22} (See for example, Eq.~(22) in Ref.~\cite{Ragone_22}). Thus, we conclude that the unitary symmetry constraints on QITE ans\"atze are the same as the ones on general variational ans\"atze. The additional constraint by QITE is considering the anti-unitary TR symmetry, which requires that only Pauli strings containing odd numbers of Pauli-$Y$ letters should be involved in the ans\"atze. 

In GQML works, one can use twirling operations~\cite{Ragone_22} to equivalently solve the linear equations (\ref{eq:master-formula-unitary-appendix}). Using twirling operations is convenient to solve the symmetry constraints of continuous groups, such as the $U(1)$ symmetry considered in subsection~\ref{sec:Example on continuous symmetry}. We use this technique to solve the $U(1)$ symmetry constraints and derive the corresponding particle number preserving Pauli strings, shown in Appendix~\ref{app:Particle-number-preserving-table}.

\section{Proof of Eq.~(\ref{eq:cTa=xia})}\label{app:cTa=xia}
We rigorously prove Eq.~(\ref{eq:cTa=xia}) according to Eq.~(\ref{eq:key-formulae}) and Eq.~(\ref{eq:mctaxiv}). We use $\ket{\cdot}$ and $\bra{\cdot}$ notation to represent column and row vectors, respectively, and gather the necessary equations from Eq.~(\ref{eq:key-formulae}) and Eq.~(\ref{eq:mctaxiv})
\begin{align}
    \left\{\begin{array}{rl}
         c^{(\bos{\alpha})}M c^{(\bos{\alpha}) T}&=M;\\
         M \ket{\bos{a}}_0&=\ket{V}; \\
         M c^{(\bos{\alpha}) T}\ket{\bos{a}}_0&=\ket{V}.
    \end{array}\right.
    \label{eq:key-formulea-app}
\end{align}
Here we focus on the unitary symmetry with $\xi=1$. We show that these equations lead to 
\begin{align}
    c^{(\bos{\alpha})} \ket{\bos{a}}_0=\ket{\bos{a}}_0,
    \label{eq:master-eq-special-solution}
\end{align}
where $\ket{\bos{a}}_0$ is derived by pseudo-inverse~\cite{kress2012numerical} according to Eq.~(\ref{eq:a0-pseudo-inverse}).

First, notice that $M$ in Eq.~(\ref{eq:M-V-calculation-measure}) is a positive and real $n\times n$ dimensional matrix~\cite{McArdle_19}, where $n$ is the dimension of the linear system of equations. it has spectrum decomposition
\begin{align}
    M=\sum_{j=1}^{n}d_j\ket{j}\bra{j},
\end{align}
with spectrum arranged by
\begin{align}
    d_1\geq \ldots \geq d_r\geq d_{r+1}=\ldots  =d_{n}=0,
\end{align}
where $r$ is the rank of $M$.  The null space $\mathcal{N}(M)\equiv\{\ket{x}\in \mathbb{C}^n:M\ket{x}=0\}=\mathrm{span}\{\ket{j},j=r+1,\ldots,n\}$ has dimension $\mathrm{dim}\mathcal{N}(M)=n-r$. The general solution of $M\ket{\bos{a}}=V$ can be decomposed into 
\begin{align}
    \ket{\bos{a}}=\ket{\bos{a}}_0+\ket{\bos{a}}_{homo},
    \label{eq:general-solution}
\end{align}
where 
\begin{align}
    \ket{\bos{a}}_0\equiv \sum_{j=1}^{r}\frac{1}{d_j}\ip{j}{V} \ket{j}
    \label{eq:special-solution-app}
\end{align} 
is the special solution given by pseudo-inverse, and $\ket{\bos{a}}_{homo}$ is the general solution of the homogeneous equation $M\ket{\bos{a}}=0$, i.e., $\ket{\bos{a}}_{homo}\in \mathcal{N}(M)$. Thus, according to $M c^{(\bos{\alpha}) T}\ket{\bos{a}}_0=\ket{V} $, $c^{(\bos{\alpha}) T}\ket{\bos{a}}_0$ belongs to the general solution Eq.~(\ref{eq:general-solution}). So the difference of $c^{(\bos{\alpha}) T}\ket{\bos{a}}_0$ and $\ket{\bos{a}}_0$ must belong to the null space, which can be expanded as
\begin{align}
    c^{(\bos{\alpha}) T}\ket{\bos{a}}_0-\ket{\bos{a}}_0=\sum_{k=r+1}^{n}\lambda_k \ket{k}.
    \label{eq:difference-expansion}
\end{align}
\begin{table*}[]
\begin{tabular}{c|cc|cc}
\# of qubits        & \multicolumn{1}{c|}{U(1) symmetric Paulis}                                                                                   & \# of Paulis & \multicolumn{1}{c|}{U(1) + TR symmetric Paulis}                                                                              & \# of Paulis \\ \hline\hline
\multirow{7}{*}{3}  & \multicolumn{1}{c|}{XYI $-$ YXI, I permutation}                                                                                & 3            & \multicolumn{1}{c|}{XYI $-$ YXI, I permutation}                                                                                & 3            \\ \cline{2-5} 
                    & \multicolumn{1}{c|}{XYZ $-$ YXZ , Z permutation}                                                                               & 3            & \multicolumn{1}{c|}{XYZ $-$ YXZ , Z permutation}                                                                               & 3            \\ \cline{2-5} 
                    & \multicolumn{1}{c|}{XXI $+$ YYI, I permutation}                                                                                & 3            & \multicolumn{2}{c}{\multirow{5}{*}{}}                                                                                                      \\ \cline{2-3}
                    & \multicolumn{1}{c|}{XXZ $+$ YYZ, Z permutation}                                                                                & 3            & \multicolumn{2}{c}{}                                                                                                                       \\ \cline{2-3}
                    & \multicolumn{1}{c|}{ZZI, I permutation}                                                                                      & 3            & \multicolumn{2}{c}{}                                                                                                                       \\ \cline{2-3}
                    & \multicolumn{1}{c|}{IIZ, Z permutation}                                                                                      & 3            & \multicolumn{2}{c}{}                                                                                                                       \\ \cline{2-3}
                    & \multicolumn{1}{c|}{ZZZ}                                                                                                     & 1            & \multicolumn{2}{c}{}                                                                                                                       \\ \hline
Total               & \multicolumn{2}{c|}{19}                                                                                                                     & \multicolumn{2}{c}{6}                                                                                                                      \\ \hline
\multirow{12}{*}{4} & \multicolumn{1}{c|}{XYII$-$YXII, II permutation}                                                                               & 6            & \multicolumn{1}{c|}{XYII$-$YXII, II permutation}                                                                               & 6            \\ \cline{2-5} 
                    & \multicolumn{1}{c|}{XYZI$-$YXZI, ZI permutation}                                                                               & 12           & \multicolumn{1}{c|}{XYZI$-$YXZI, ZI permutation}                                                                               & 12           \\ \cline{2-5} 
                    & \multicolumn{1}{c|}{XYZZ$-$YXZZ, ZZ permutation}                                                                               & 6            & \multicolumn{1}{c|}{XYZZ$-$YXZZ, ZZ permutation}                                                                               & 6            \\ \cline{2-5} 
                    & \multicolumn{1}{c|}{\begin{tabular}[c]{@{}c@{}}XXXY$-$YYYX$+$XYYY$-$YXXX\\ XXYX$-$YYXY$+$XYYY$-$YXXX\\ XYXX$-$YXYY$+$XYYY$-$YXXX\end{tabular}} & 3            & \multicolumn{1}{c|}{\begin{tabular}[c]{@{}c@{}}XXXY$-$YYYX$+$XYYY$-$YXXX\\ XXYX$-$YYXY$+$XYYY$-$YXXX\\ XYXX$-$YXYY$+$XYYY$-$YXXX\end{tabular}} & 3            \\ \cline{2-5} 
                    & \multicolumn{1}{c|}{XXII$+$YYII, II permutation}                                                                               & 6            & \multicolumn{2}{c}{\multirow{8}{*}{}}                                                                                                      \\ \cline{2-3}
                    & \multicolumn{1}{c|}{XXZI$+$YYZI, ZI permutation}                                                                               & 12           & \multicolumn{2}{c}{}                                                                                                                       \\ \cline{2-3}
                    & \multicolumn{1}{c|}{XXZZ$+$YYZZ, ZZ permutation}                                                                               & 6            & \multicolumn{2}{c}{}                                                                                                                       \\ \cline{2-3}
                    & \multicolumn{1}{c|}{\begin{tabular}[c]{@{}c@{}}XXYY$+$YYXX$+$XXXX$+$YYYY\\ XYXY$+$YXYX$+$XXXX$+$YYYY\\ XYYX$+$YXXY$+$XXXX$+$YYYY\end{tabular}} & 3            & \multicolumn{2}{c}{}                                                                                                                       \\ \cline{2-3}
                    & \multicolumn{1}{c|}{IIIZ,  Z permutation}                                                                                    & 4            & \multicolumn{2}{c}{}                                                                                                                       \\ \cline{2-3}
                    & \multicolumn{1}{c|}{IIZZ, ZZ permutation}                                                                                    & 6            & \multicolumn{2}{c}{}                                                                                                                       \\ \cline{2-3}
                    & \multicolumn{1}{c|}{ZZZI, I permutation}                                                                                     & 4            & \multicolumn{2}{c}{}                                                                                                                       \\ \cline{2-3}
                    & \multicolumn{1}{c|}{ZZZZ}                                                                                                    & 1            & \multicolumn{2}{c}{}                                                                                                                       \\ \hline
Total               & \multicolumn{2}{c|}{69}                                                                                                                     & \multicolumn{2}{c}{27}                                                                                                                     \\ \hline
\end{tabular}
\caption{$U(1)$ symmetry preserving Pauli strings on 3 and 4 qubits. Pauli strings in the left column preserve $U(1)$ symmetry. In the right column, each Pauli string has an odd number of Pauli-$Y$ letters, preserving both $U(1)$ and TR symmetry. The total number of Pauli strings for the corresponding number of qubits is presented in the row ``Total".}
\label{tab:u1-symmetry-more-qubits}
\end{table*}

In the following proof, we will show that $\lambda_k=0$. According to Eq.~(\ref{eq:difference-expansion}), $\lambda_k$ can be calculated by 
\begin{align}
    \lambda_k=\bra{k}c^{(\bos{\alpha}) T}\ket{\bos{a}}_0-\ip{k}{\bos{a}}_0, k=r+1,\dots,n.
\end{align}
Notice that the special solution $\ket{\bos{a}}_0$ is orthogonal to the null space, thus $\ip{k}{\bos{a}}_0=0$. Taking the explicit expression of pseudo-inverse Eq.~(\ref{eq:special-solution-app}) leads to 
\begin{align}
    \lambda_k=\sum_{j=1}^{r}\frac{1}{d_j}\ip{j}{V} \bra{k}c^{(\bos{\alpha}) T}\ket{j}.
    \label{eq:lambda}
\end{align}

Now we only need to show $\bra{k}c^{(\bos{\alpha}) T}\ket{j}=0$. By applying both side of $c^{(\bos{\alpha}) T}M c^{(\bos{\alpha})}=M$ (Equivalent to the first equation in Eq.~(\ref{eq:key-formulea-app})) on a null-space vector $\ket{k}$, one finds
\begin{align}
    \sum_{j=1}^{r} d_j c^{(\bos{\alpha})T}\ket{j}\melem{j}{c^{(\bos{\alpha})}}{k}=0,
\end{align}
where the right-hand side vanishes due to the definition of the null space. Since $c^{(\bos{\alpha})T}\ket{j}$ are linear independent vectors, we have $\melem{j}{c^{(\bos{\alpha})}}{k}=0$. Thus $ \lambda_k=0, \forall k=1,\ldots, r$ according to Eq.~(\ref{eq:lambda}).

Finally, we find that the right-hand side of Eq.(\ref{eq:difference-expansion}) is zero, and Eq.(\ref{eq:master-eq-special-solution}) is proved according to the real orthogonality of $c^{(\bos{\alpha})}$. $\qed$

\section{Particle number preserving Pauli strings}\label{app:Particle-number-preserving-table}

In Table~\ref{tab:u1-symmetry-more-qubits}, we present particle number preserving Pauli strings on 3 and 4 qubits. Pauli strings with the same permutation structure are written in one block for conciseness. For example, \textit{$XYI-YXI$, $I$ permutation} in the first line of the 3-qubit case denotes three Pauli strings: $XYI-YXI$, $XIY-YIX$ and $IXY-IYX$, where the letter $I$ permutes on three qubits. Then the Pauli exponentials to be added in the ans\"atze are
\begin{align}
    e^{i\theta_1(XYI-YXI)},e^{i\theta_2(XIY-YIX)}, e^{i\theta_3(IXY-IYX)},
\end{align}
with free parameters $\theta_1$, $\theta_2$ and $\theta_3$ evolved using VarQITE.

Table~\ref{tab:u1-symmetry-more-qubits} is an extension of the 2-qubit case in Table~\ref{table:u1-symmetry}. Using Pauli strings in this table, one can design the particle number preserving ansatz for the quantum chemistry model with Hamiltonian $H_\mathrm{chem}$ defined in Eq.~(\ref{eq:quantum-chemistry-model}). For example, for a local interaction term in $H_\mathrm{chem}$ acting on four qubits, one should add all the $69$ Pauli exponentials to the ans\"atze (See the last line of the table). If the local interaction term also preserves TR symmetry, the number of Pauli exponentials is reduced to $27$.

\bibliographystyle{unsrt}
\bibliography{main}
\end{document}